\documentclass[times,twocolumn,final]{elsarticle}

\usepackage{framed,multirow}
\usepackage{tabularx}

\usepackage{amssymb}
\usepackage{latexsym}

\usepackage{url}
\usepackage{xcolor}
\usepackage{hyperref}
\usepackage{amsmath}
\usepackage{colortbl}
\usepackage{float}
\usepackage{threeparttable}
\usepackage{booktabs}
\usepackage{pifont}
\newcommand{\cmark}{\ding{51}}%
\newcommand{\xmark}{\ding{55}}%
\usepackage{bm}

\usepackage{microtype}
\usepackage[modulo, switch]{lineno}

\def\r{{\mathbf r}}

\definecolor{MyDarkGreen}{rgb}{0.0, 0.65, 0.0}
\newcommand{\RNum}[1]{\uppercase\expandafter{\romannumeral #1\relax}}
\newcommand{\pStar}{\phantom{*}}
\newcommand{\bStar}{\textbf{*}}

\definecolor{shadecolor}{rgb}{0.9, 0.9, 0.9}
\definecolor{aero}{rgb}{0.78515625, 1.0, 0.898039215686275}
\definecolor{faintGreen}{rgb}{0.8, 1.0, 0.8}
\definecolor{lowteal}{RGB}{180, 230, 225}    
\definecolor{highgold}{RGB}{255, 220, 150}   

\definecolor{newcolor}{rgb}{.8,.349,.1}

\begin{document}



\title{UNISELF: A Unified Network with Instance Normalization and Self-Ensembled Lesion Fusion for Multiple Sclerosis Lesion Segmentation}

\author[1]{Jinwei Zhang\thanks{Corresponding author: jwzhang@jhu.edu}}
\author[2]{Lianrui Zuo}
\author[3]{Blake E. Dewey}
\author[4]{Samuel W. Remedios}
\author[2]{Yihao Liu}
\author[1]{Savannah P. Hays}
\author[5]{Dzung L. Pham}
\author[3]{Ellen M. Mowry}
\author[3]{Scott D. Newsome}
\author[3]{Peter A. Calabresi}
\author[1]{Aaron Carass}
\author[1]{Jerry L. Prince}

\address[1]{Department of Electrical and Computer Engineering, Johns Hopkins University, Baltimore, MD 21218, USA}
\address[2]{Department of Electrical and Computer Engineering, Vanderbilt University, Nashville, TN 37215, USA}
\address[3]{Department of Neurology, Johns Hopkins School of Medicine, Baltimore, MD 21287, USA}
\address[4]{Department of Computer Science, Johns Hopkins University, Baltimore, MD 21218, USA}
\address[5]{Department of Radiology, Uniformed Services University of the Health Sciences, Bethesda, MD, 20814, USA}


\begin{abstract}
Automated segmentation of multiple sclerosis~(MS) lesions using multicontrast magnetic resonance~(MR) images improves efficiency and reproducibility compared to manual delineation, with deep learning (DL) methods achieving state-of-the-art performance.
However, these DL-based methods have yet to simultaneously optimize in-domain accuracy and out-of-domain generalization when trained on a single source with limited data, or their performance has been unsatisfactory.
To fill this gap, we propose a method called UNISELF, which achieves high accuracy within a single training domain while demonstrating strong generalizability across multiple out-of-domain test datasets. 
UNISELF employs a novel test-time self-ensembled lesion fusion to improve segmentation accuracy, and leverages test-time instance normalization (TTIN) of latent features to address domain shifts and missing input contrasts.
Trained on the ISBI 2015 longitudinal MS segmentation challenge training dataset, UNISELF ranks among the best-performing methods on the challenge test dataset.
Additionally, UNISELF outperforms all benchmark methods trained on the same ISBI training data across diverse out-of-domain test datasets with domain shifts and missing contrasts, including the public MICCAI 2016 and UMCL datasets, as well as a private multisite dataset. 
These test datasets exhibit domain shifts and/or missing contrasts caused by variations in acquisition protocols, scanner types, and imaging artifacts arising from imperfect acquisition.
\textcolor{black}{Our code is available at \url{https://github.com/uponacceptance}.}
\end{abstract}


\maketitle


\section{Introduction}
\label{sec:intro}

Multiple sclerosis~(MS) is characterized by chronic inflammatory demyelination and neurodegeneration of the central nervous system~\citep{haider2016topograpy}.
Magnetic resonance imaging~(MRI) is commonly used for diagnosing and monitoring MS in clinics due to its sensitivity to focal tissue inflammation~\citep{filippi2019assessment}.
In MS, demyelinating white matter lesions typically appear hyperintense on T2-weighted~(T2w) and T2-weighted Fluid-Attenuated Inversion Recovery~(FLAIR) images, and hypointense on T1-weighted~(T1w) images.
Advancements in MRI, such as higher spatial resolution, improved signal-to-noise ratio, and biomarkers specific to the pathology of MS disease (e.g. iron rim lesions~\citep{dal2017slow}) make analyzing the longitudinal evolution of the progression of the lesion more informative. 
These advancements improve the reliability of treatment evaluations~\citep{kolb2022pathology}.
\textcolor{black}{Accurate lesion mask delineations serve as a foundational step for quantifying lesion volumes and extracting advanced biomarkers relevant to lesion progression, such as iron rim lesions, which are linked to chronic active inflammation in MS~\citep{kaunzner2019quantitative}.}
However, manual delineation of MS lesions faces challenges in efficiency and reproducibility due to their heterogeneous shape, size, and location (as shown in Fig. \ref{fig:fig3}).
\textcolor{black}{In addition, Inter-rater variability~\citep{carass2017longitudinal, carass2020sr, commowick2021ni, styner2008challenge, zijdenbos1994tmi} and lack of consistency across imaging sites and protocols~\citep{carass2024nir} make manual segmentation particularly difficult, especially in multi-center and retrospective studies.}
To address these challenges, researchers have focused on developing automated lesion segmentation methods for many years.
\textcolor{black}{However, variations in image quality across scanners~\citep{van2024scanner} and the limited availability of manually delineated datasets for training learning-based methods~\citep{carass2017dib} pose challenges for automated segmentation methods.}

Automated lesion segmentation methods can be categorized as classical or learning-based depending on whether image features are hand-crafted or learned from data.
For classical methods, early approaches included the use of hand-crafted image features such as atlas-based topology priors with an additional lesion class~\citep{shiee2010topology} and probabilistic models for lesion classification and growth~\citep{schmidt2012automated}, \textcolor{black}{as well as more recent probabilistic segmentation models like SAMSEG~\citep{cerri2021contrast}.}
For learning-based methods, early feature extraction techniques have been used, such as sparse coding of image patches through dictionary learning~\citep{weiss2013multiple}.
With the emergence of deep learning, recent research has primarily involved training convolutional neural network models using manually delineated labels to learn image features.
Pioneering work includes 3D patch-based cascaded architecture~\citep{valverde2017improving}, 2D slice-based multi-branch architecture~\citep{aslani2019multi}, and 3D volume-based encoder-decoder architecture~\citep{brosch2016deep}.
Since then, deep learning methods have demonstrated improved performance over earlier methods in terms of both accuracy and efficiency~\textcolor{black}{\citep{jiang2023review, ma2022multiple}, with state-of-the-art models such as LST-AI~\citep{wiltgen2024lst} and HD-MS-Lesions~\citep{brugnara2020automated} being publicly released along with their trained weights.} 

Some recent work has focused on improving segmentation accuracy when trained with limited data with manual delineations by adapting the widely used U-Net~\citep{ronneberger2015u} architecture. 
U-Net has the advantages of both multiscale feature representation and skip connections, enabling accurate MS lesion segmentation.
For example, the best-performing published lesion segmentation methods evaluated using the 2015 ISBI longitudinal MS lesion segmentation challenge data~\citep{carass2017longitudinal, carass2017dib} are Tiramisu~\citep{zhang2019multiple} and ALL-Net~\citep{zhang2021all}.
In Tiramisu, the convolutional layer in the U-Net was replaced with a densely connected convolutional layer~\citep{jegou2017one} to improve feature learning. 
In ALL-Net, a coordinate convolutional layer was incorporated into their U-Net to capture anatomical information.
Other modifications in network architecture include the addition of attention mechanisms in U-Net to improve segmentation accuracy, such as slice-wise~\citep{zhang2019rsanet} and folded attention U-Net for 3D volumes~\citep{zhang2021efficient}, U-Net with squeeze and attention modules~\citep{rondinella2023boosting}, and attention-gated U-Net~\citep{hashemi2022delve}.
Those variants of U-Net have been reported to improve segmentation accuracy for in-domain tests, such as the ISBI challenge, when the data distributions between training and test are identical.
\textcolor{black}{In parallel, the nnU-Net framework~\citep{isensee2021nnu} emerged as a self-configuring U-Net-based architecture that has become a widely adopted state-of-the-art baseline across various medical image segmentation tasks, including MS lesion segmentation.
For example, nnU-Net has demonstrated strong performance in more recent MS lesion segmentation challenges, such as in the MS new lesion segmentation challenge MSSEG-2~\citep{commowick2021msseg}.}

In addition to improving accuracy on in-domain tests when trained with limited data, it is equally important to deal with out-of-domain shifts when the test data distribution deviates from the training.
Domain shifts in MRI for MS lesions include variations in image contrast due to scanner or protocol differences, as well as imaging artifacts introduced during acquisition.
Those domain shifts can potentially cause generalization errors~\citep{quinonero2022dataset} and reduce the accuracy of trained segmentation models.
There are three philosophies for addressing domain shifts for MS lesion segmentation: domain harmonization, adaptation, and generalization.
For domain harmonization, the variation in contrast between scanners is mitigated by using synthesis-based harmonization techniques~\citep{roy2010isbi, dewey2019deepharmony, zuo2022disentangling, zuo2023haca3, gebre2023cross}.
Harmonized images are expected to maintain consistent contrast in both training and test data thus prevent domain shifts; \citet{carass2024nir}~demonstrated that this is at least true with respect to manually generated delineations.
For domain adaptation, pre-trained models are adapted to a target test domain using new labeled data from that domain.
Example adaptation methods include one-shot adaptation~\citep{valverde2019one} and harmonization-enriched domain adaptation~\citep{zhang2023harmonization}.
\textcolor{black}{Additionally, unsupervised domain adaptation can be performed without requiring target domain labels~\citep{gerin2024exploring}.}
For domain generalization, invariant/adaptive features are imposed to reduce domain shifts.
Methods such as spatially adaptive sub-networks~\citep{kamraoui2022deeplesionbrain}, domain-invariant latent features~\citep{zhao2021robust, aslani2020scanner}, contrast-adaptive modeling~\citep{cerri2021contrast}, domain generalization augmentation~\citep{zhang2023domain}, and federated learning~\citep{liu2022ms} have been proposed, contributing to more reliable automated MS lesion segmentation in clinical and multisite settings.
\textcolor{black}{Recent efforts have further emphasized the importance of addressing real-world domain shifts through approaches such as test-time training~\citep{gerin2024exploring}, domain randomization~\citep{billot2023synthseg}, and benchmarking on distribution shifts~\citep{malinin2022shifts}, with methods including deep ensembles~\citep{lakshminarayanan2017simple}.}

The handling of varying sets of available contrast-weighted MR images across different datasets is a crucial generalization ability for MS lesion segmentation in clinical practice.  
For example, the 2015 ISBI challenge~\citep{carass2017longitudinal} dataset provides T1w, T2w, proton-density-weighted~(PDw), and FLAIR images with every subject, but it is not common for all four of these contrasts to be available in clinical practice. 
Most prior MS segmentation methods assume that the same set of contrasts is available during both training and testing and fix the network input accordingly, which limits their flexibility for a broader clinical deployment. 
Only a few studies have explored the ability of networks to handle missing contrasts.
\citet{feng2019self} applied a contrast dropout training strategy by randomly replacing a subset of all available contrasts with constant values (e.g., all zeros) when forming the training inputs without changing the network architecture.
ModDrop++~\citep{liu2022moddrop++} improved contrast dropout performance using a dynamic network architecture and a co-training loss involving full and missing contrasts during training.
In~\citet{zhang2023domain}, contrast dropout was combined with domain generalization augmentation to further improve generalization performance and combat missing contrasts.

Despite the success of deep learning methods for MS segmentation, few have been shown to simultaneously achieve strong in-domain accuracy, robust out-of-domain generalization, and the ability to handle missing contrasts.
Furthermore, their performance has not been satisfactory, especially with limited single-source training data. 
For example, Tiramisu~\citep{zhang2019multiple} and ALL-Net~\citep{zhang2021all} achieved the highest accuracy in the ISBI challenge, but used fixed multicontrast inputs and did not validate multisite generalization.
DeepLesionBrain~\citep{kamraoui2022deeplesionbrain} demonstrated good generalization to unseen domains, but it had lower in-domain accuracy than Tiramisu and used fixed network inputs.
Data augmentation with contrast dropout~\citep{zhang2023domain} also exhibited good generalization, but its in-domain accuracy was not validated.
ModDrop++~\citep{liu2022moddrop++} trained a unified model capable of handling various input contrasts while maintaining accuracy compared to independently trained models, but its generalization to unseen domains was unknown.
Image harmonization methods such as HACA3~\citep{zuo2023haca3} have the potential to improve generalization in downstream lesion segmentation by harmonizing multicontrast images and imputing missing contrasts. 
However, the generalization of HACA3 itself should be guaranteed before its application, and training a generalizable HACA3 model requires a multisite training dataset, which may not be readily available to all institutions.

To address the aforementioned gap, we propose a method called UNISELF~(\textbf{U}nified \textbf{N}etwork with \textbf{I}nstance normalization and \textbf{S}elf-\textbf{E}nsembled \textbf{L}esion \textbf{F}usion) which \textbf{trains a model with limited single source training data that can simultaneously achieve high segmentation accuracy and generalization}. 
To train the model, we used the publicly available single-site 2015 ISBI challenge training dataset~\citep{carass2017longitudinal}, which contains only 5 subjects with an average of 4.2 longitudinal scans per subject in training. 
Our results demonstrate superior in-domain accuracy on the ISBI challenge test dataset, as well as strong generalization to domain shifts and missing contrasts across multiple multisite test datasets.
This work extends our conference paper~\citep{zhang2023towards} and includes the following contributions:
\begin{enumerate}
    \item UNISELF includes a novel self-ensembled lesion fusion strategy that augments multi-orientation MRI inputs and ensembles the augmented outputs using a two-step lesion detection and growth approach to improve accuracy;
    \item UNISELF leverages test-time instance normalization (TTIN) that normalizes latent features for each test input to improve model generalization in handling domain shifts and varying MRI contrasts;
    \item The generalization ability of UNISELF was evaluated using extensive heterogeneous multisite test data, demonstrating its superior performance compared to other benchmarks.
\end{enumerate}
\textcolor{black}{Our code is available at \url{https://github.com/uponacceptance}.}

\section{Related works}
\label{sec:related_works}

\subsection{Ensemble Learning}
Ensemble learning is a technique that combines multiple diverse learning algorithms or models to capture a wider range of unique patterns and features compared to individual algorithms~\citep{opitz1999popular, kuncheva2003measures}, thus enhancing robustness and accuracy.
In MS lesion segmentation, prior works such as \cite{tohidi2022multiple} trained multiple models with diverse architectures to achieve better results compared to individual models. 
However, the performance of the ensemble in \cite{tohidi2022multiple} was not compared to benchmark methods and datasets like Tiramisu~\citep{zhang2019multiple} and ALL-Net~\citep{zhang2021all} in the 2015 ISBI challenge.
Furthermore, selecting architectures to improve model diversity can be a laborious and arbitrary process, sometimes even resulting in worse performance compared to dedicatedly designed single models~\citep{zhang2021all}.

\subsection{Test-time Augmentation}
Data augmentation addresses the overfitting of trained models by enhancing the size and variety of limited training data by applying multiple transformations such as flips and rotations to each sample during training~\citep{shorten2019survey}. 
Test-time augmentation (TTA) applies multiple transformations to test data in the same way as during training, thus generating a more robust prediction from multiple predictions compared to a single prediction without augmentation~\citep{shorten2019survey}.
TTA resembles ensemble learning in the data space and has been used in medical image segmentation~\citep{moshkov2020test, wang2019aleatoric, henschel2020fastsurfer, isensee2021nnu}.
For MS lesion segmentation, a widely used TTA approach generates three 3D lesion masks, each derived from one of the three cardinal planes (axial, sagittal, and coronal), followed by voxel-wise majority voting to obtain the final 3D mask~\citep{zhang2019multiple}.
However, this aggregation strategy does not consider the multi-instance nature of MS lesion detection and segmentation.
Moreover, comprehensive multi-orientation augmentations including flipping and rotation are not considered in this approach.

\subsection{Latent Feature Regulation}
Batch normalization (BN)~\citep{ioffe2015batch} has been widely used in MS lesion segmentation models under the assumption that there are no domain shifts between training and test data~\citep{zhang2019multiple, zhang2021all}.
However, BN stores domain-specific training feature statistics to normalize test-time features, which may not generalize well to out-of-domain test datasets.   
To improve out-of-domain generalization, prior works in generalizable MS lesion segmentation have proposed strategies to generate domain-invariant features regardless of site changes~\citep{zhao2021robust, aslani2020scanner}.
However, these works require multisite training data with lesion mask labels, which may be difficult to obtain.

\subsection{\textcolor{black}{Handling Missing Contrast}}
\textcolor{black}{Handling missing MRI contrasts is critical for deploying segmentation models in real-world clinical environments, where not all contrast-weighted MR images are consistently available.
Early work by \citet{feng2019self} proposed contrast dropout, which trains the network with randomly dropping input contrasts~(replacing them with zeros) to improve robustness to missing contrasts at inference.
This strategy was extended in ModDrop++~\citep{liu2022moddrop++}, which introduced a dynamic network architecture and a co-training loss between full and partial contrast inputs to learn contrast-invariant features.
\citet{zhang2023domain} further enhanced generalization by combining contrast dropout with domain generalization augmentation.
However, these approaches either do not explicitly address the intrinsic distribution shifts caused by varying input contrasts or require complex training procedures.}

\section{Method}

We propose UNISELF to address the limitations of existing test-time augmentation (data ensemble learning), latent feature regulation, and missing contrast handling methods in improving MS lesion segmentation accuracy and generalization. 
These limitations are further compounded by the challenges of limited training data and variations in MRI image quality and available contrasts.
UNISELF overcomes these limitations with a segmentation model that simultaneously achieves high in-domain accuracy, strong out-of-domain generalization, and robustness to varying contrasts, even when trained on limited single-site data.
The two key building blocks of UNISELF, self-ensembled lesion fusion and test-time instance normalization, are detailed in Sections~\ref{ssec:self} and~\ref{ssec:ttin}, respectively.

\begin{figure}[!t]

\begin{minipage}[b]{1.0\linewidth}
  \centering
  \centerline{\includegraphics[width=8.5cm]{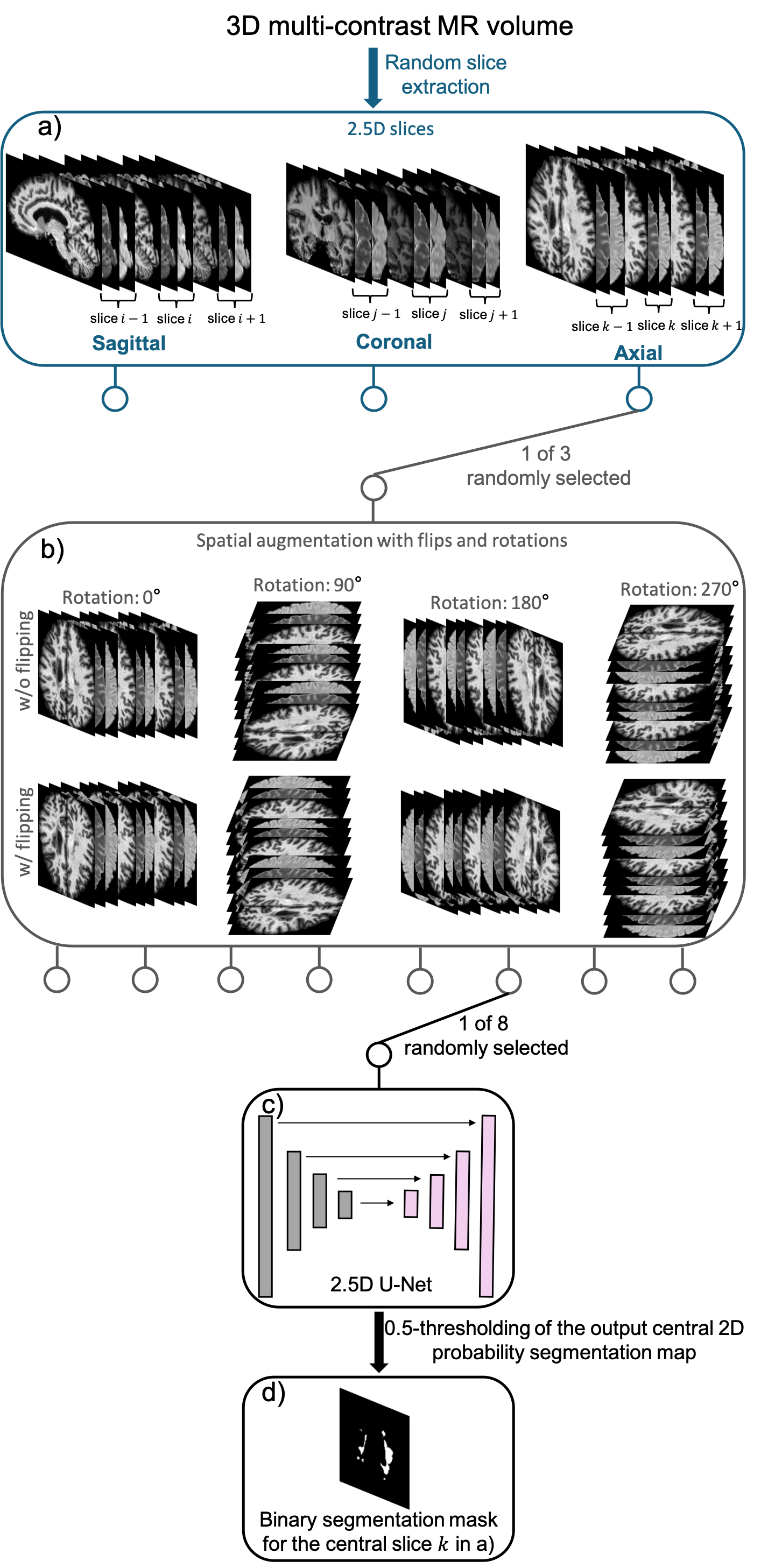}}
\end{minipage}
\caption{An illustration of the spatial augmentation, network input, and network output during training in UNISELF.}
\label{fig:fig1}
\end{figure}

\subsection{Self-Ensembled Lesion Fusion}
\label{ssec:self}
In UNISELF, we introduce a novel self-ensembled lesion fusion strategy to improve model accuracy. 
\textcolor{black}{Unlike existing TTA approaches in medical image segmentation such as those used in FastSurfer~\citep{henschel2020fastsurfer} and nnU-Net~\citep{isensee2021nnu}, which aggregate multi-orientation TTA predictions via~(weighted) averaging, \textbf{we propose a novel two-step lesion detection and growth approach to fuse multi-orientation TTA predictions}.
Specifically, the first step, ``lesion detection'', identifies lesioned voxels with high confidence. 
The second step, ``lesion growth'', expands these detected lesions to neighboring voxels to recover the full lesion extent.
Together, these two steps form our ``lesion fusion'' strategy, which aggregates multi-orientation TTA predictions into a final segmentation using a single~(unified) model.
This lesion fusion strategy forms the first major methodological contribution of our work, as detailed in the following sections.}

\subsubsection*{Multi-orientation Augmentation in Training}
In UNISELF, a single U-Net is trained to slice-wise process multicontrast 3D MRI volumes with multi-orientation augmentation.
Figure~\ref{fig:fig1} presents an overview of the multi-orientation data augmentation process applied during training. 
In Fig.~\ref{fig:fig1}a, three adjacent 2D slices of multicontrast MRI, concatenated along the channel dimension with zero-padding at the boundaries (referred to as 2.5D input~\citep{zhang2019multiple}), are extracted from a random selection of the three cardinal planes: axial, sagittal, or coronal. 
In Fig.~\ref{fig:fig1}b, a randomly selected augmentation from 8 possible augmentations, including flips (vertical and horizontal) and rotations ($0^{\circ}$, $90^{\circ}$, $180^{\circ}$, and $270^{\circ}$), is applied to the extracted 2.5D input. 
These augmentations alter the spatial configuration of the 2.5D input without changing its intensity values, preserving the original image information while enhancing the diversity of inputs encountered by the network during training.
In Fig.~\ref{fig:fig1}c, the augmented 2.5D input is fed into the network to generate the probabilistic output map of the central 2D slice.
In Fig.~\ref{fig:fig1}d, a binary segmentation mask is obtained from the augmented 2.5D input by applying a threshold of 0.5 to the output probabilistic map, classifying pixels with values above 0.5 as lesion and those below as non-lesion areas.
The ground truth lesion mask of the central slice undergoes the same augmentation process as in Figs.~\ref{fig:fig1}a and b and is used in the loss function for backpropagation.

\subsubsection*{Confidence Map in Test Time}
\begin{figure}[!t]

\begin{minipage}[b]{1.0\linewidth}
  \centering
\centerline{\includegraphics[width=8.5cm]{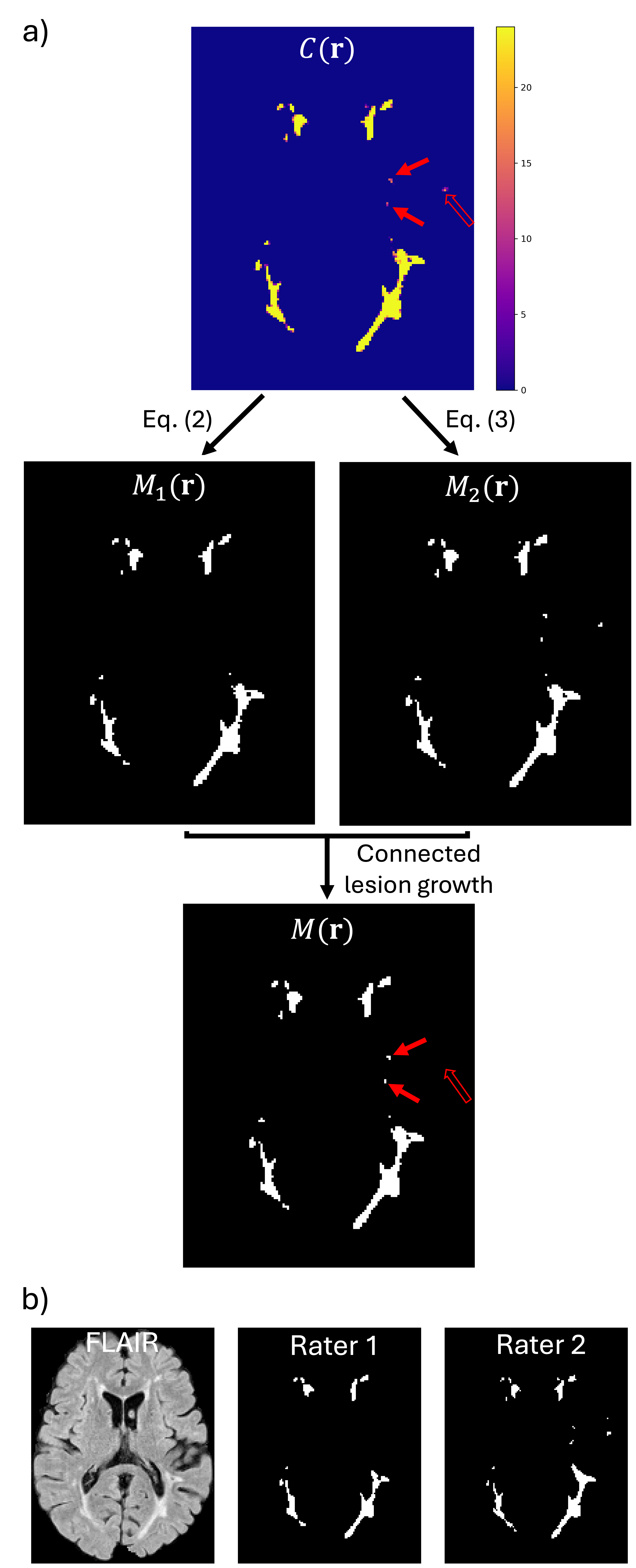}}
\end{minipage}
\caption{
Illustration of the proposed self-ensembled lesion fusion strategy in UNISELF. 
(a):~Confidence map $C(\r)$ (top), two binary lesion masks $M_1(\r)$ and $M_2(\r)$ (middle), and final binary lesion mask $M(\r)$ (bottom).
(b):~Corresponding FLAIR image with lesion delineations from two raters as reference.
}
\label{fig:fig2}
\end{figure}

At test time, a \emph{confidence map} is generated through comprehensive multi-orientation processing and aggregation of the same multicontrast 3D volume using the trained segmentation model.
Specifically, the eight spatial augmentations shown in Fig.~\ref{fig:fig1}b are applied to each of the three cardinal plane 3D volumes, resulting in $N_M = 8\times3=24$ augmented multicontrast 3D volumes.
For any augmented input volume with index $i$~$(1 \leq i \leq N_M)$, the trained network processes the volume in a 2.5D fashion, where 2D binary segmentation masks are generated slice-by-slice and subsequently stacked to form a 3D binary mask corresponding to the augmented input volume.
The resulting 3D binary mask is then flipped and rotated back into the original space, denoted as $M_i(\r)$, where $\r$ represents the spatial coordinate of a 3D voxel in the original space.
A confidence map $C(\r)$ with integer values between $0$ and $N_M$ is then generated by adding $N_M$ 3D binary masks in the original space: 
\begin{equation}
\label{confidence_map}
C(\r) = \sum_{i=1}^{N_M} M_i(\r), \ \forall \r.
\end{equation}
This confidence map indicates the number of times that the network, trained with all cardinal planes, rotations, and flips, predicts a voxel as part of a lesion.
Figure~\ref{fig:fig2}a shows an example $C(\r)$ as a heatmap ranging between integer values of 0 and 24.
For example, $C(\r)=24$ means that the model consistently predicts the voxel at $\r$ as belonging to a lesion under all spatial augmentations.
We use $C(\r)$ to derive the final lesion segmentation mask through a two-step process of lesion detection and connected lesion growth, as detailed in the next section.

\subsubsection*{Lesion Detection and Connected Growth}
\label{sssec:voting}

After obtaining $C(\r)$ using Eq.~\ref{confidence_map}, the next step applies two thresholds $\tau_1$ and $\tau_2$, where $\tau_1 \geq \tau_2$, on $C(\r)$ to generate two binary segmentation masks: 
\begin{equation}
\label{voting1}
M_1(\r) = \left\{ \begin{array}{ll} 1 & C(\r) > \tau_1, 
\\ 0 & \text{otherwise} \end{array} \right. 
\ \forall \r, 
\end{equation}
and
\begin{equation}
\label{voting2}
M_2(\r) = \left\{ \begin{array}{ll} 1 & C(\r) > \tau_2, 
\\ 0 & \text{otherwise} \end{array} \right. 
\ \forall \r.
\end{equation}
Figure~\ref{fig:fig2}a shows an example of $M_1(\r)$ and $M_2(\r)$.
Because $\tau_1 \ge \tau_2$, the binary region defined by $M_2(\r)$ always includes that defined by $M_1(\r)$. 
Moreover, since the voxels in $M_1(\r)$ have greater certainty of belonging to a lesion, they are considered to belong to lesions without further evaluation and are referred to as \emph{detected lesions}. 

We use $M_2(\r)$ in a \emph{connected lesion growth} step to provide candidates to also include within the detected lesions. 
The connected lesion growth step produces the final segmentation $M(\mathbf{r})$ by iteratively expanding the detected lesions defined by $M_1(\mathbf{r})$ to include additional spatially connected voxels identified in $M_2(\mathbf{r})$.
We use 26-connectivity for this growth process.   
Figure~\ref{fig:fig2}a shows an example of the final segmentation $M(\r)$ after connected lesion growth, where segmentation masks of small lesions (denoted by solid red arrows, present in other axial slices in $M_1(\r)$ that are not shown in Fig.~\ref{fig:fig2}a) are grown and refined in $M(\r)$, while lesions not detected in $M_1(\r)$ (denoted by hollow red arrows) do not appear in $M(\r)$ either.
Figure~\ref{fig:fig2}b shows the corresponding FLAIR image with lesion masks delineated by two raters as reference.
We experimentally determine optimal values for $\tau_1$ and $\tau_2$~(see Section~\ref{tau12}) to ensure their effectiveness in the proposed self-ensemble strategy.

\subsection{Test-time Instance Normalization (TTIN)}
\label{ssec:ttin}

To improve UNISELF generalization, we proposed to leverage TTIN to address potential latent feature mismatches \textcolor{black}{caused by (1)~out-of-domain shifts and (2)~varying input contrast:
(1)~Specifically, we demonstrate that TTIN mitigates feature statistics shifts caused by out-of-domain shifts;
(2)~Most importantly, we observe that varying combinations of input contrasts also induces shifts in feature statistics, even when such combinations were present via contrast dropout during training. 
\textbf{To mitigate this, we proposed the use of TTIN to recalibrate instance normalization statistics at inference time on a per-input basis, preserving invariant feature representations across contrast variations.}
Integrating TTIN with contrast dropout forms the second major methodological contribution of our work, as explained in the following sections.}

\subsubsection*{Batch Normalization in Training and Testing}
The distribution of latent features in any layer can be regulated by Batch normalization (BN) to stabilize and accelerate stochastic gradient descent optimization~\citep{ioffe2015batch}.
During mini-batch training, for a layer with feature map $\bm{x} = (x_1, x_2, \cdots, x_c) \in \mathbb{R}^{n \times c}$, where $n$ denotes the batch size, and $c$ denotes the feature/channel dimension, BN transforms $\bm{x}$ in a feature-wise manner with the following steps:
\begin{subequations}
\label{normalization}
\begin{align}
    \hat{x}_j &= \frac{x_j - \mathbb{E}[x_j]}{\sqrt{\text{Var}[x_j] + \epsilon}} \label{normalization_a}, \\
    y_j &= \gamma_j\hat{x}_j + \beta_j \label{normalization_b}
\end{align}
\end{subequations}
where $\gamma_j$ and $\beta_j$ in Eq.~\ref{normalization_b} are learnable parameters for each feature/channel $j \in \{1, \cdots, c\}$.
During BN training, the moving averages of $\mathbb{E}[x_j]$ and $\text{Var}[x_j]$ over all mini-batches are stored and denoted as $\mathbb{E}_{MA}[x_j]$ and $\text{Var}_{MA}[x_j]$, respectively.
At test time, $\mathbb{E}[x_j]$ and $\text{Var}[x_j]$ in Eq.~\ref{normalization_a} are replaced with $\mathbb{E}_{MA}[x_j]$ and $\text{Var}_{MA}[x_j]$ for all test data, and $\gamma_j$ and $\beta_j$ in Eq.~\ref{normalization_b} are retained from training~\citep{ioffe2015batch}.

\subsubsection*{Domain Shifts in Feature Space}
In test scenarios with domain shifts such as contrast variation or missing contrast in multisite MRI data, a mismatch in latent features between training and test datasets may occur.
A visualization of such latent space feature mismatch normalized by $\mathbb{E}_{MA}[x_j]$ and $\text{Var}_{MA}[x_j]$ is presented in Section~\ref{ttin_vis} with a detailed experimental setup.
Distinct clusters corresponding to each test site are observed in the latent space features of both shallow and deep layers in a U-Net. 
This indicates a mismatch across different test sites resulting from domain shifts in the test data distribution relative to the training data distribution (i.e., $\mathbb{E}_{MA}[x_j]$ and $\text{Var}_{MA}[x_j]$).

\subsubsection*{TTIN to Handle Domain Shifts}

Given the potential mismatch between training and test domain features normalized with $\mathbb{E}_{MA}[x_j]$ and $\text{Var}_{MA}[x_j]$, we propose to use \emph{instance-specific} statistics, i.e., test-time instance normalization (TTIN), for any given input at test time to handle domain shifts.
Specifically, $\mathbb{E}[x_j]$ and $\text{Var}[x_j]$ in Eq.~\ref{normalization_a} are computed for each test input (per 2.5D slice, \textcolor{black}{using a batch size of one during inference}), instead of using $\mathbb{E}_{MA}[x_j]$ and $\text{Var}_{MA}[x_j]$ from training.
The parameters $\gamma_j$ and $\beta_j$ in Eq.~\ref{normalization_b} are retained from the training phase.
In Section~\ref{ttin_vis} we demonstrate that the features normalized by TTIN appear mixed without distinct clusters, indicating improved alignment across various test sites.

Since $\gamma_j$ and $\beta_j$ are retained from training, feature normalization methods on the training of $\gamma_j$ and $\beta_j$ in Eq.~\eqref{normalization_b} may affect the performance of TTIN.
Different feature normalization methods for training $\gamma_j$ and $\beta_j$ in Eq.~\eqref{normalization_b} and their impact on TTIN is assessed in the experiments section~(see Sections~\ref{ttin_indomin} and~\ref{ttin_outofdomin}), including BN, IN, and conditional IN (CondIN)~\citep{dumoulin2016learned}, where CondIN involves conditional $\gamma_j$ and $\beta_j$ for each input contrast combination.


\subsection{Implementation Details}
\label{subsec:training_details}
We use U-Net~\citep{ronneberger2015u} as the backbone architecture for UNISELF.
\textcolor{black}{Specifically, we adopt a 5-level encoder-decoder U-Net with 64, 128, 256, 512, and 1024 feature channels at each level, respectively. 
Each level contains two sequential blocks of 3$\times$3 convolution, ReLU activation, and feature normalization. 
Downsampling is performed using 2$\times$2 max pooling, while upsampling is performed using nearest-neighbor interpolation followed by a convolution.}
Training details are listed below:
\begin{itemize}
    \item Prior to the multi-orientation augmentation in Fig.~\ref{fig:fig1}, 3D elastic or affine transformations were randomly applied using TorchIO~\citep{perez2021torchio}~(with a probability of 75\% for each multicontrast 3D volume) to augment brain tissue and lesion shapes.
    \item \textcolor{black}{For each training batch, individual samples were drawn by first randomly selecting a subject and then randomly extracting a 2.5D slice that contained at least one lesion voxel in the central slice. 
    All samples in the batch were taken from the same imaging plane (e.g., axial) and underwent identical spatial augmentations, including random rotation and flipping.}
    \item The $L_2$ loss was used between the predicted probabilistic segmentation map and the label for back-propagation, as it demonstrated better overall performance compared to the Dice and focal losses~\citep{zhang2019multiple}.
    \item Random sampling of two-rater labels was employed~\citep{jungo2018effect}.
    \item Contrast dropout (CD)~\citep{feng2019self} was applied by randomly replacing a subset of available contrasts in each training batch with all-zero images.
    \item The Adam optimizer~\citep{kingma2014adam} was used to update the network weights with an initial learning rate of $10^{-4}$, a mini-batch size of 12, and 150 epochs with 300 iterations per epoch.
\end{itemize}

\section{Experiments and Results}
\label{sec:results}

\subsection{Datasets}
\begin{figure*}[!t]

\begin{minipage}[b]{1.0\linewidth}
  \centering
  \centerline{\includegraphics[width=18cm]{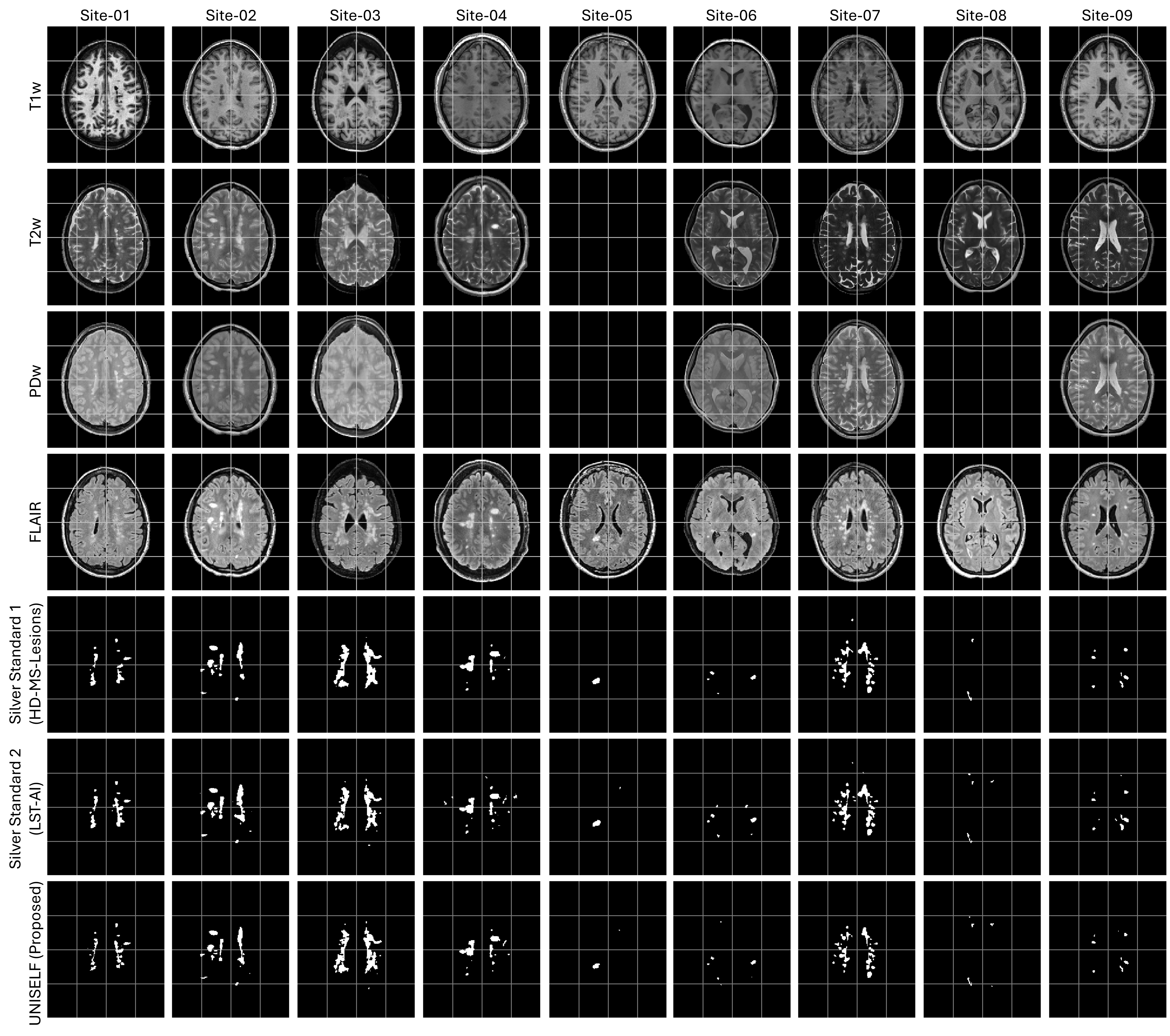}}
\end{minipage}
\caption{
Example images from the private multisite test dataset with silver standard lesion segmentation labels and UNISELF predictions. 
}
\label{fig:fig3}
\end{figure*}

\subsubsection{Training Dataset: Public 2015 ISBI Challenge}
The 2015 ISBI longitudinal MS segmentation challenge~\citep{carass2017longitudinal} training dataset was used to train all models in this work.
This training dataset consists of 21 time points from 5 subjects acquired using a single scanner. 
Four contrasts (T1w, T2w, PDw, and FLAIR) with two expert delineations are available for all time points.
All data were rigidly registered to the $1 \text{mm}^3$ MNI space using MIPAV’s optimized automatic registration\footnote{\url{https://mipav.cit.nih.gov/}} followed by brain extraction and dura removal\footnote{\url{https://www.nitrc.org/projects/toads- cruise/}}, as well as N4 bias field correction\footnote{\url{https://stnava.github.io/ANTs/}}.

\subsubsection{Test Dataset~\RNum{1}: Public 2015 ISBI Challenge}
The 2015 ISBI challenge test dataset consists of 61 time points from 14 subjects acquired using the same scanner as the training dataset, with the same four contrasts and pre-processing steps~\citep{carass2017longitudinal}.
Expert delineations are not available to the public.
Instead, the test performance is assessed by submitting the segmentation results to the challenge website to obtain a score.

\subsubsection{Test Dataset~\RNum{2}: Public 2016 MICCAI Challenge}
The 2016 MICCAI MS segmentation challenge~\citep{commowick2018objective} contains a training dataset of 15 subjects acquired from three different scanners.
Five contrasts (T1w, T1w with gadolinium, T2w, PDw, and FLAIR) with a consensus of seven expert delineations are available for all subjects.
For each subject, each contrast was denoised with the non-local means algorithm~\citep{coupe2008optimized}, followed by rigid body registration to the corresponding FLAIR contrast, skull stripping, and N4 bias field correction.
Rigid registration to the ICBM 152 MNI space was applied to each contrast.
T1w images with gadolinium contrast were not used to ensure that the input multicontrast images at test time are a subset of those used in training.

\subsubsection{Test Dataset~\RNum{3}: Public UMCL Data}
The UMCL~(University Medical Center Ljubljana) dataset~\citep{lesjak2018novel} contains 30 subjects acquired using a single scanner.
Four contrasts (T1w, T1w with gadolinium, T2w, and FLAIR) with a consensus of three expert delineations are available for all subjects.
All data were registered to the $1 \text{mm}^3$ MNI space with N4 correction.
T1w images with gadolinium contrast were not used.

\subsubsection{Test Dataset~\RNum{4}: Private Multisite Data}

In our study, a private multisite data set was used for additional generalization validation.
This multisite dataset contains 93 subjects acquired from \textcolor{black}{9} scanners (sites) \textcolor{black}{in a clinical setting}, with various available T1w, T2w, PDw, and FLAIR contrasts, but does not include manual delineations of lesions.
\textcolor{black}{Among the 93 subjects, the acquisition resolution statistics \(([ \text{frequency},\ \text{phase},\ \text{slice encodings} ])\) for each MRI contrast are summarized as follows:
\begin{itemize}
  \item \textbf{T1w}: 92 3D scans $([1.02\pm0.08,\ 1.02\pm0.08,\ 1.04\pm0.08]~\text{mm}^3)$, 1 2D scan $([1.07,\ 0.75,\ 5.00]~\text{mm}^3)$, and 0 missing.
  \item \textbf{T2w}: 16 3D scans $([0.98\pm0.03,\ 0.98\pm0.03,\ 1.00\pm0.00]~\text{mm}^3)$, 72 2D scans $([0.86\pm0.19,\ 0.80\pm0.23,\ 3.22\pm0.85]~\text{mm}^3)$, and 5 missing.
  \item \textbf{PDw}: 0 3D scans, 63 2D scans $([0.89\pm0.16,\ 0.85\pm0.23,\ 2.87\pm0.51]~\text{mm}^3)$, and 30 missing.
  \item \textbf{FLAIR}: 78 3D scans $([1.01\pm0.06,\ 0.99\pm0.03,\ 1.02\pm0.08]~\text{mm}^3)$, 13 2D scans $([0.88\pm0.11,\ 0.88\pm0.06,\ 3.38\pm1.07]~\text{mm}^3)$, and 2 missing.
\end{itemize}
}
All data went through similar pre-processing steps as the ISBI data, with an additional super-resolution step using SMORE~\citep{remedios2023self, zhao2020smore} on the multislice 2D acquired images before registration to the $1 \text{mm}^3$ MNI space.
Representative multicontrast images after pre-processing are shown in rows 1--4 of Fig.~\ref{fig:fig3}.

\begin{figure*}[!t]

\begin{minipage}[b]{1.0\linewidth}
  \centering
  \centerline{\includegraphics[width=18cm]{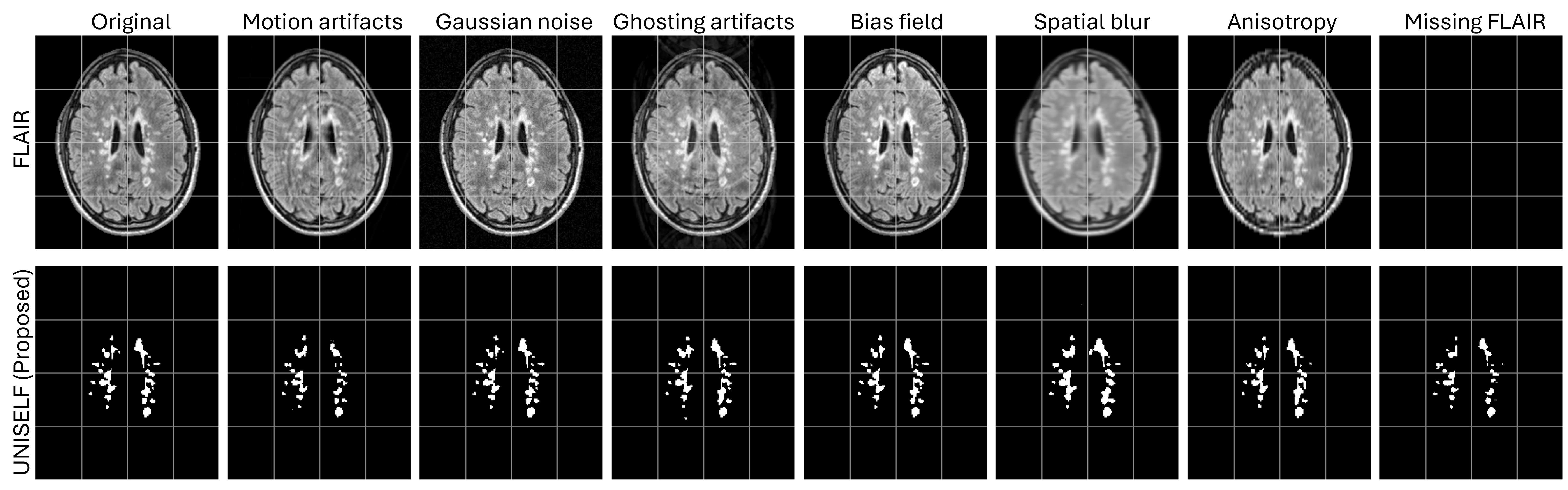}}
\end{minipage}
\caption{
Example FLAIR images of `Site-07' in Fig.~\ref{fig:fig3} with various simulated artifacts and the corresponding UNISELF predictions. 
}
\label{fig:fig4}
\end{figure*}

To further validate the generalization performance of trained models to imaging artifacts, FLAIR images in the Test Dataset~\RNum{4} were corrupted with various artifacts, including motion, Gaussian noise, ghosting, bias field, spatial blur, and anisotropy, using the TorchIO data augmentation library~\citep{perez2021torchio}, as well as missing FLAIR.
Examples of corrupted images of one representative subject are shown in Fig.~\ref{fig:fig4}.

\subsubsection{Silver Standard Delineation Generation on Test Dataset~\RNum{4}}
Since manual lesion delineations are not available for the Test Dataset~\RNum{4}, we have designed the following protocol to generate silver standard delineations: 
\textcolor{black}{
    \begin{itemize}
    \item First, we employed HACA3~\citep{zuo2023haca3} to impute missing contrasts for each subject in Test Dataset~\RNum{4}.
    To ensure the generalization ability of HACA3, we trained the HACA3 model on a collection of the ISBI training dataset and Test Dataset~\RNum{4} before applying it for contrast imputation.
    \item Second, we adopted two independent, publicly available models---LST-AI~\citep{wiltgen2024lst} and HD-MS-Lesions~\citep{brugnara2020automated}---to generate two silver standard delineations on Test Dataset~\RNum{4}~(examples shown in rows~5 and~6 of Fig.~\ref{fig:fig3}) after missing contrast imputation.
    These models were trained on large, heterogeneous MS datasets in their respective studies and require fixed input combinations: ``T1w+FLAIR'' for LST-AI and ``T1w+T2w+FLAIR'' for HD-MS-Lesions.
    \item Finally, we evaluated the segmentation performance of both the benchmark methods and UNISELF by averaging the segmentation scores~(as defined in Eq.~\eqref{score}) across the two independent silver standard delineations, in accordance with the ISBI challenge protocol~\citep{carass2017longitudinal}, which relies on annotations from two human raters.
\end{itemize}
We note that missing contrast imputation for Test Dataset~\RNum{4} was applied only within the silver standard delineation generation protocol.
In all subsequent test experiments, all methods were evaluated using the original set of available multicontrast images for each subject in Test Dataset~\RNum{4}.
}


\begin{table*}[!t]
\centering
\caption{
Performance comparison and ablation study of the self-ensembled lesion fusion strategy~(Section~\ref{ssec:self}) in UNISELF on the 2015 ISBI Challenge test data.
\textcolor{black}{``UNISELF~(Proposed)'' refers to the finalized UNISELF configuration selected through cross-validation, as described in Section~\ref{cross_val}.
Average performance metrics for ALL-Net and Tiramisu were obtained directly from their respective publications, as subject-level results were not publicly available.}
}
\label{tab:table1}
\scalebox{0.70}{
\begin{tabular}{l ccccccc}
\toprule
\multicolumn{1}{c}{\textbf{Method}} & \textbf{Score} ($\uparrow$) & \textbf{DSC} ($\uparrow$) & \textbf{PPV} ($\uparrow$) & \textbf{TPR} ($\uparrow$) & \textbf{LFPR} ($\downarrow$) & \textbf{LTPR} ($\uparrow$) & \textbf{VC} ($\uparrow$)\\
\cmidrule(lr){1-1}
\cmidrule(lr){2-8}
ALL-Net~\citep{zhang2021all} & \bf{93.32} & 0.639 & \underline{0.914} & 0.525 & \underline{0.122} & \underline{0.533} & 0.860 \\
Tiramisu~\citep{zhang2019multiple} & 93.21 & 0.643 & 0.908 & 0.533 & 0.124 & 0.520 & \bf{0.867} \\
UNISELF (Proposed) & \textcolor{black}{\underline{93.28 ± 7.40}} & \textcolor{black}{\bf{0.664 ± 0.141}} & \textcolor{black}{0.882 ± 0.099} & \textcolor{black}{\bf{0.569 ± 0.199}} & \textcolor{black}{\bf{0.100 ± 0.102}} & \textcolor{black}{0.508 ± 0.230} & \textcolor{black}{\underline{0.866}} \\
\cmidrule(lr){1-1}
\cmidrule(lr){2-8}
\multicolumn{1}{c}{\textbf{Ablation of self-ensembled lesion fusion }} \\
\cmidrule(lr){1-1}
\cmidrule(lr){2-8}
Removing connected lesion growth & \textcolor{black}{\pStar{}93.12 ± 7.57\bStar{}} & \textcolor{black}{\pStar{}0.627 ± 0.158\bStar{}} & \textcolor{black}{\pStar{}\bf{0.922 ± 0.073}\bStar{}} & \textcolor{black}{\pStar{}0.508 ± 0.197\bStar{}} & \textcolor{black}{\pStar{}\underline{0.118 ± 0.109}\bStar{}} & \textcolor{black}{\pStar{}0.502 ± 0.227\bStar{}} & \textcolor{black}{\underline{0.866}} \\
Removing lesion detection and connected lesion growth & 
\textcolor{black}{\pStar{}93.01 ± 7.41\bStar{}} & \textcolor{black}{\pStar{}\underline{0.645 ± 0.148}\bStar{}} & \textcolor{black}{\pStar{}0.900 ± 0.088\bStar{}} & \textcolor{black}{\pStar{}\underline{0.536 ± 0.196}\bStar{}} & \textcolor{black}{\pStar{}0.164 ± 0.124\bStar{}} & \textcolor{black}{\pStar{}\bf{0.535 ± 0.232}\bStar{}} & \textcolor{black}{\bf{0.867}} \\
\bottomrule
\end{tabular}} \\
{\footnotesize (\textcolor{black}{\bStar{}: statistically significant difference compared to UNISELF (Proposed) in each column, based on the paired Wilcoxon signed-rank test ($p < 0.05$).}
The best and second-best performances in each column are denoted in \textbf{bold} and \underline{underline}, respectively.)}
\end{table*}

\subsection{Benchmarks and Performance Evaluation}
\label{benchmarks}
For out-of-domain generalization tests, we included benchmarks Tiramisu~\citep{zhang2019multiple}, ModDrop++~\citep{liu2022moddrop++}, CD~\citep{feng2019self}, DG augmentation using both BN and IN~\citep{zhang2023domain}.
These benchmarks were trained using the same training strategies described in Section~\ref{subsec:training_details}, based on their publicly available network architecture code, \textcolor{black}{with customized adaptations applied to specific methods.}
\textcolor{black}{For training, we used the same 3D affine and elastic transformation, training data sampling, loss function, random label sampling, and optimizer across methods, to ensure a controlled comparison focused solely on the impact of architectural differences of these benchmarks.
During inference, we used a batch size of one for all methods, including those relying on BN statistics and those using TTIN.
Customized adaptations for each benchmark include:
\begin{itemize}
    \item CD training was applied to both CD~\citep{feng2019self} and ModDrop++~\citep{liu2022moddrop++} baselines.
    \item Independent models, each corresponding to a fixed subset of input contrasts, were trained for Tiramisu~\citep{zhang2019multiple}.
    \item A co-training loss involving full and missing contrasts as proposed in ModDrop++~\citep{liu2022moddrop++} was applied during the training of ModDrop++.
    \item Domain generalization (DG) augmentation was applied during the training of the DG baseline using BN and IN~\citep{zhang2023domain}, which included contrast dropout, imaging artifacts (motion, Gaussian noise, ghosting, bias field, spatial blur, and anisotropy) using TorchIO~\citep{perez2021torchio}, and gamma transformation to simulate contrast variation.
    This forms a conservative comparison against the DG baseline, which was exposed to artifact corruptions during training.
    \item Random sampling of 3D patches (patch size: $112 \times 112 \times 112$) during training and sliding-window inference (step size: $56 \times 56 \times 56$) were implemented for the CD baseline~\citep{feng2019self}, which employed a 3D U-Net architecture.
\end{itemize}
}
\textcolor{black}{We also benchmarked against nnU-Net~\citep{isensee2021nnu}, following the recommended training commands provided in the official nnU-Net repository to ensure a standardized setup and reproducibility. 
Since CD is not supported in nnU-Net, we trained independent nnU-Net models, each corresponding to a fixed subset of input contrasts.}
Other potential benchmarks such as ALL-Net~\citep{zhang2021all} and DeepLesionBrain~\citep{kamraoui2022deeplesionbrain} were not included due to the unavailability of their source code.

Evaluation measures are the same as those used in the ISBI 2015 challenge~\citep{carass2017longitudinal}: Dice Similarity Coefficient~(DSC), Precision~(PPV), Sensitivity~(TPR), Lesion-wise True Positive Rate~(LTPR), Lesion-wise False Positive Rate~(LFPR), Pearson’s correlation coefficient of the lesion volumes~(VC), and a score by a weighted average of the above:
\begin{equation}
\label{score}
\text{Score} = \frac{\text{DSC}}{8} + \frac{\text{PPV}}{8} + \frac{\text{LTPR}}{4} + \frac{1 - \text{LFPR}}{4} + \frac{\text{VC}}{4}.
\end{equation}
The scores of the 2015 ISBI challenge test were obtained by submitting segmentation results and receiving normalized scores computed by the challenge website~\citep{carass2017longitudinal}.
The scores of the other test datasets were computed using Eq.~\eqref{score}.
\textcolor{black}{We employed the Wilcoxon signed-rank test~\citep{rey2011wilcoxon}, a non-parametric test for paired samples, to assess the statistical significance of performance differences.
To account for multiple comparisons, we applied the Benjamini–Hochberg procedure~\citep{hochberg1990more} to control the false discovery rate at a significance level of 0.05.}

\subsection{\textcolor{black}{Cross-validation to Finalize UNISELF Configuration}}
\label{cross_val}
\textcolor{black}{To determine the final UNISELF configuration prior to testing---including the parameters $\tau_1$ and $\tau_2$ used in self-ensembled lesion fusion (Section~\ref{ssec:self}), and the choice of normalization technique (i.e., BN, IN, or CondIN) used to train $\gamma_j$ and $\beta_j$ in Eq.~4b for TTIN (Section~\ref{ssec:ttin})---we performed five-fold cross-validation on the ISBI challenge training set, holding out one subject out of five in each fold for validation.
The validation criterion was the segmentation score defined in Eq.~\eqref{score}.
Figure~\hyperlink{figS1}{S1} shows the cross-validation scores from a grid search of $\tau_1$ and $\tau_2$ for models trained with BN (Fig.~\hyperlink{figS1}{S1}a), IN (Fig.~\hyperlink{figS1}{S1}b), and CondIN (Fig.~\hyperlink{figS1}{S1}c).
The best validation performance was achieved with $\tau_1 = 16$ and $\tau_2 = 7$ using CondIN training, and this configuration was fixed for all subsequent test experiments.}

\subsection{Comparison in the ISBI Test}

Table~\ref{tab:table1} shows a comparison of UNISELF with ALL-Net~\citep{zhang2021all} and Tiramisu~\citep{zhang2019multiple} as two of the best-performing methods in the ISBI challenge. 
The comparison is based on the metrics reported in the respective papers, and the finalized UNISELF configuration is determined through cross-validation, as described in Section~\ref{cross_val}.
UNISELF achieved a comparable Score to ALL-Net and Tiramisu, with lower PPV and LTPR, but higher DSC, TPR, and (1-LFPR).

An ablation study was conducted on the ISBI challenge test dataset by progressively removing the connected lesion growth and lesion detection steps as described in Section~\ref{sssec:voting}.
When both steps were removed, the network's output was simplified to a majority-vote segmentation derived from three cardinal planes without flips or rotations.
As presented in Table~\ref{tab:table1}, improved performance in terms of Score was observed by progressively adding the lesion detection and connected lesion growth steps.

\textcolor{black}{In terms of inference speed, UNISELF requires 2 minutes and 30 seconds with a GPU memory usage of 656~MB on an NVIDIA Quadro RTX 5000.}

\subsection{Comparison in Out-of-Domain Tests}

\begin{figure*}[!t]

\begin{minipage}[b]{1.0\linewidth}
  \centering
  \centerline{\includegraphics[width=18cm]{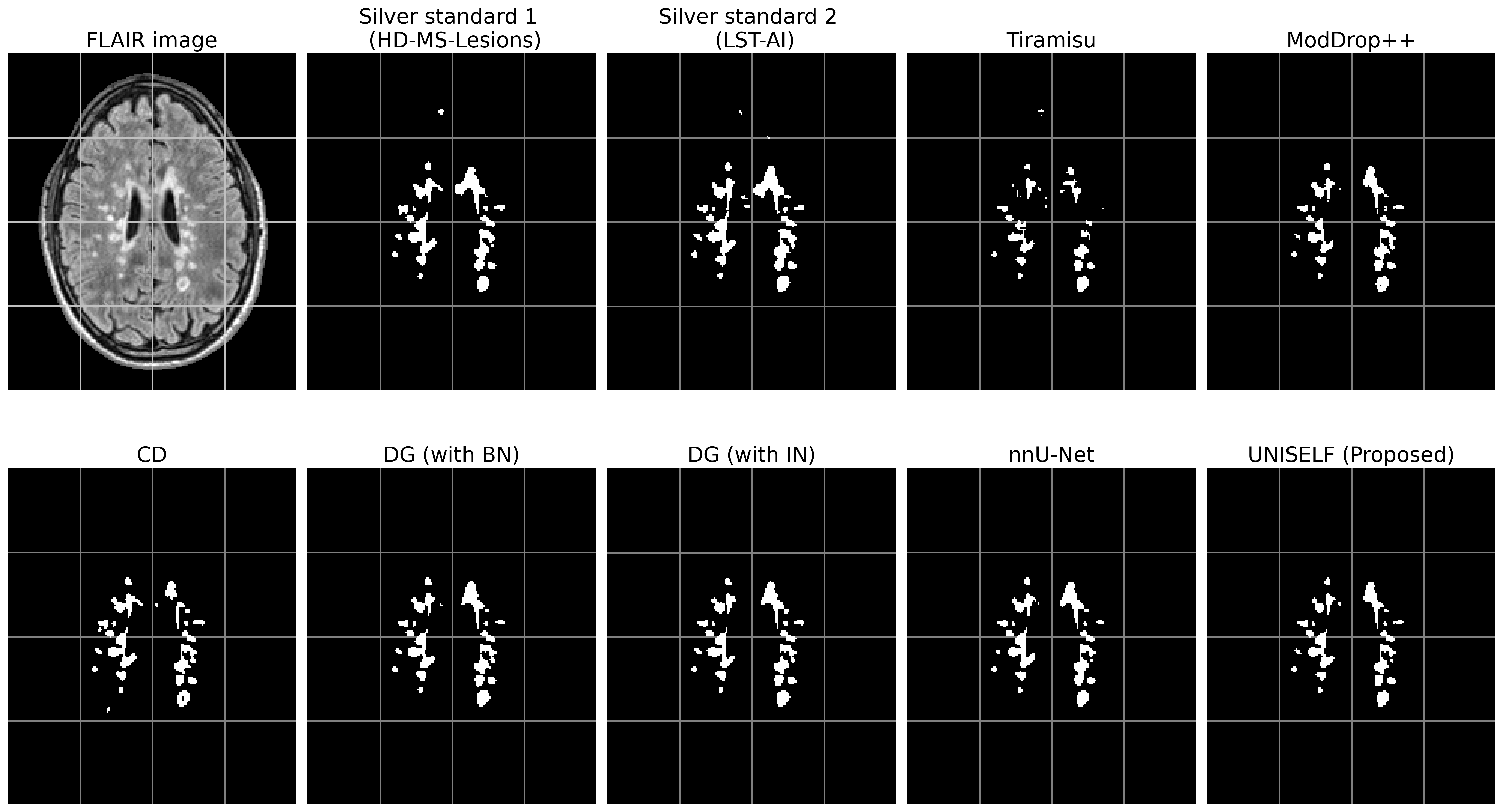}}
\end{minipage}
\caption{
MS segmentation masks from different methods on a representative subject in the private test dataset.
``UNISELF~(Proposed)'' refers to the finalized UNISELF configuration selected through cross-validation, as described in Section~\ref{cross_val}.
}
\label{fig:fig5}
\end{figure*}

\begin{figure*}[!t]

\begin{minipage}[b]{1.0\linewidth}
  \centering
  \centerline{\includegraphics[width=18cm]{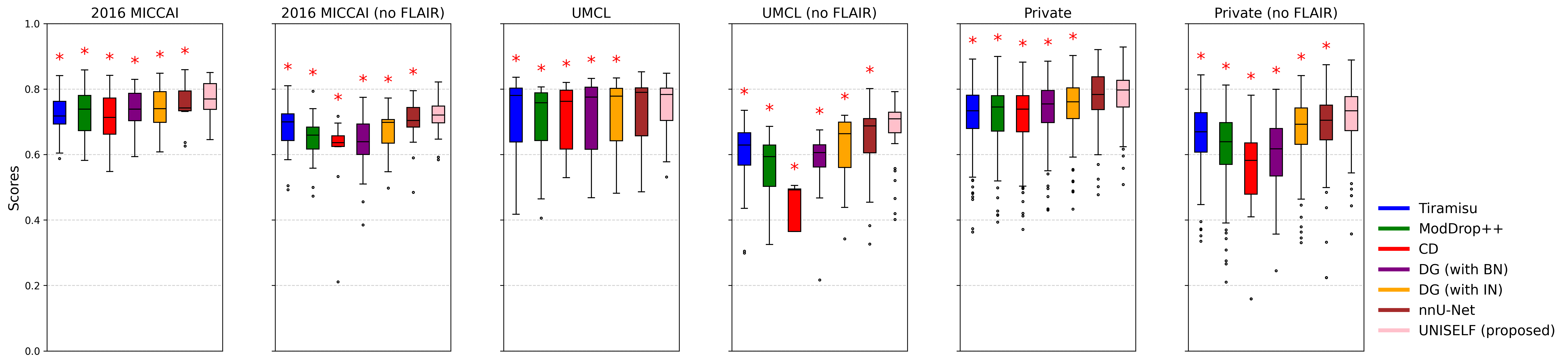}}
\end{minipage}
\caption{Segmentation scores~(Eq.~\eqref{score}) by different methods on public and private test datasets considering both original and missing FLAIR multicontrast inputs.
``UNISELF~(Proposed)'' refers to the finalized UNISELF configuration selected through cross-validation, as described in Section~\ref{cross_val}. 
(Red star: statistically significant difference compared to UNISELF (Proposed) in each boxplot, based on the paired Wilcoxon signed-rank test, $p < 0.05$.)
}
\label{fig:fig6}
\end{figure*}

\begin{figure*}[!t]

\begin{minipage}[b]{1.0\linewidth}
  \centering
  \centerline{\includegraphics[width=18cm]{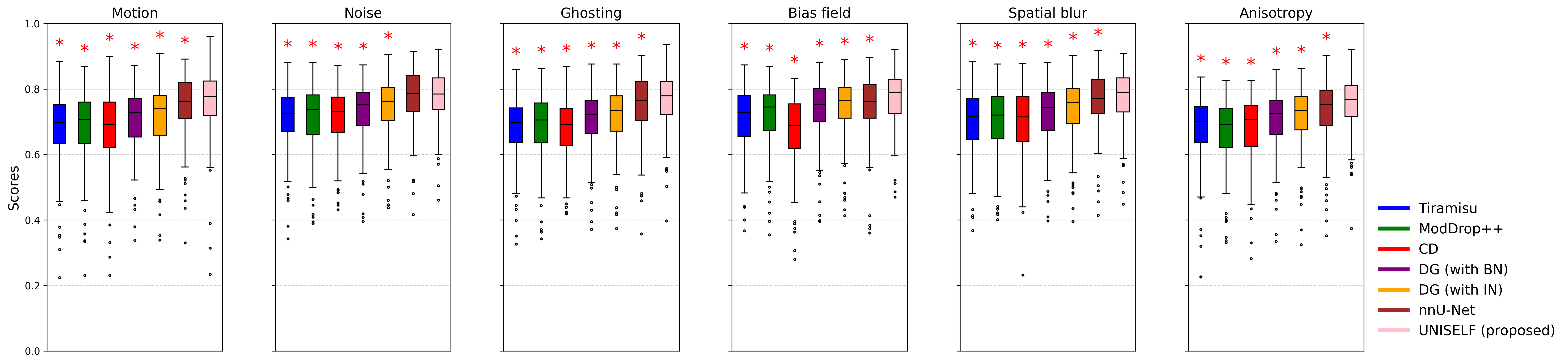}}
\end{minipage}
\caption{Segmentation scores~(Eq.~\eqref{score}) by different methods on the private multisite test dataset with various FLAIR artifacts.
``UNISELF~(Proposed)'' refers to the finalized UNISELF configuration selected through cross-validation, as described in Section~\ref{cross_val}. 
(Red star: statistically significant difference compared to UNISELF (Proposed) in each boxplot, based on the paired Wilcoxon signed-rank test, $p < 0.05$.)
}
\label{fig:fig7}
\end{figure*}

Figures~\ref{fig:fig5},~\hyperlink{figS2}{S2}, and~\hyperlink{figS3}{S3} show MS lesion masks predicted by different methods on a representative subject from the private test dataset, demonstrating segmentation on the original contrast~(Fig.~\ref{fig:fig5}), FLAIR with motion artifact~(Fig.~\hyperlink{figS2}{S2}), and input with missing FLAIR~(Fig.~\hyperlink{figS3}{S3}).
In Fig.~\ref{fig:fig5}, missed lesions were observed in the benchmark methods Tiramisu and CD~(contrast dropout). 
In contrast, the benchmark methods DG~(domain generalization) with BN and IN, \textcolor{black}{the benchmark method nnU-Net,} and the proposed method UNISELF showed very few missed lesions.
In Fig.~\hyperlink{figS2}{S2}, with motion artifacts added to the FLAIR image, more missed lesions were observed in the benchmark methods Tiramisu, ModDrop++, and CD,  but were not observed in DG with BN and IN, \textcolor{black}{nnU-Net,} and UNISELF.
In Fig.~\hyperlink{figS3}{S3}, only UNISELF and the benchmark method DG with IN demonstrated robust segmentation performance when dealing with missing FLAIR.

Figure~\ref{fig:fig6} shows boxplots of segmentation scores~(Eq.~\eqref{score}) across Test Datasets~\RNum{2}, \RNum{3}, and~\RNum{4} for the original multicontrast inputs and the corresponding inputs without FLAIR.
UNISELF achieved noticeable improvements compared to the benchmark methods, especially when the FLAIR contrast was missing. 
Figure~\ref{fig:fig7} shows boxplots of segmentation scores~(Eq.~\eqref{score}) across Test Dataset~\RNum{4} with various artifacts added to the FLAIR contrast.
\textcolor{black}{The benchmark methods DG with IN and BN had already been exposed to such artifacts during training through DG augmentation.
Nevertheless, UNISELF outperformed all benchmark methods, including DG with BN and IN.}

\subsection{\textcolor{black}{Ablation Study}}
\textcolor{black}{
We conducted an ablation study on the two key components from Sections~\ref{ssec:self} and~\ref{ssec:ttin}, using both 2.5D (as implemented in the proposed UNISELF) and 3D U-Net architectures.
The study systematically examined the impact of removing these components: (1)~replacing self-ensembled lesion fusion~(Section~\ref{ssec:self}) with conventional three-plane majority voting, and (2)~replacing TTIN~(Section~\ref{ssec:ttin}) with BN training statistics.
The 3D U-Net architecture includes patch-based sliding-window processing, as implemented in nnU-Net~\citep{isensee2021nnu}, using a patch size of $112 \times 112 \times 112$ and a step size of $56 \times 56 \times 56$.
Figures~\hyperlink{figS4}{S4} and~\hyperlink{figS5}{S5} show boxplots of segmentation scores~(Eq.~\eqref{score}) from the ablation study, corresponding to the same out-of-domain experiments shown in Figs.~\ref{fig:fig6} and~\ref{fig:fig7}.
For both 3D and 2.5D U-Net architectures, adding self-ensembled lesion fusion and TTIN progressively improved their segmentation scores, especially in Fig.~\hyperlink{figS5}{S5}, which shows results on Test Dataset IV with various artifacts added to the FLAIR contrast.
Moreover, the 2.5D U-Net~(as implemented in the proposed UNISELF) generally outperformed the 3D U-Net across all out-of-domain experiments.}

\subsection{Visualization of latent features normalized by TTIN}
\label{ttin_vis}

\begin{figure}[!t]

\begin{minipage}[b]{1.0\linewidth}
  \centering
  \centerline{\includegraphics[width=8.5cm]{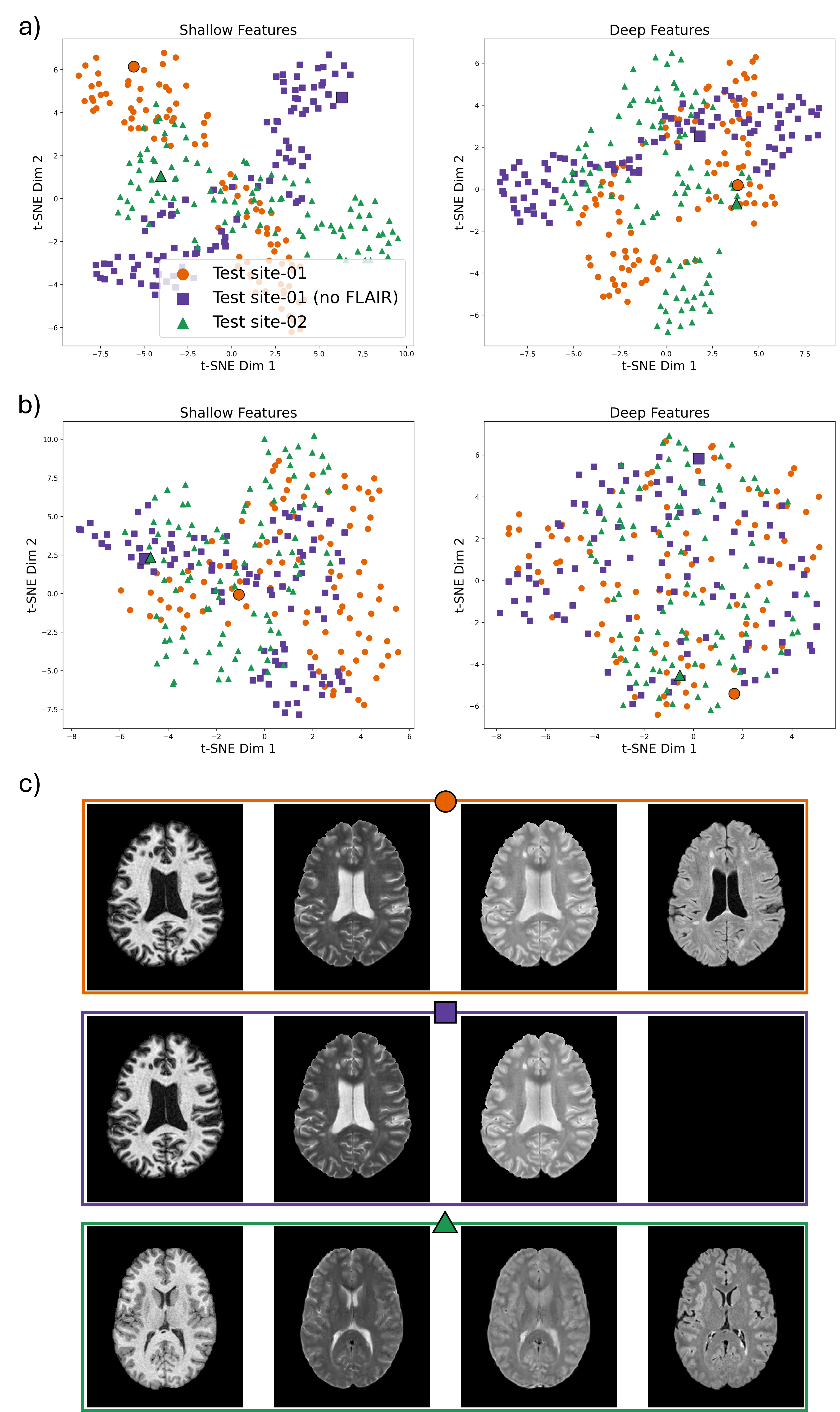}}
\end{minipage}
\caption{
t-SNE visualization of feature statistics across three test datasets.
(a):~t-SNE of normalized features in shallow~(left) and deep~(right) layer features using training dataset normalization statistics. 
(b):~Corresponding t-SNE using instance-specific normalization statistics at test time as proposed in UNISELF. 
(c):~Example multicontrast images (represented by larger markers in~(a) and~(b)) from the three test datasets.
}
\label{fig:fig8}
\end{figure}

Figure~\ref{fig:fig8} presents t-SNE plots of U-Net's (trained with CD and BN) shallow~(highest level in the encoder path) and deep~(lowest level in the encoder path) layer features across three test sites.
In this visualization, t-SNE was applied to the concatenated channel-wise mean and variance of the normalized features.
The three test sites include: ISBI challenge test dataset~(`Site-01'), a test site from Test Dataset~\RNum{4}~(`Site-02'), and a modified version of the ISBI challenge test dataset where the FLAIR contrast was replaced with zeros~(`Site-01 (no FLAIR)').
Ten axial slices, spaced 10 slices apart, were extracted from each of the 10 subjects at each test site, with each dot in Fig.~\ref{fig:fig8} representing one of these inputs.
Figure~\ref{fig:fig8}a shows t-SNE plots of features normalized using the training dataset normalization statistics $\mathbb{E}_{MA}[x_j]$ and $\text{Var}_{MA}[x_j]$ of the corresponding layer~\citep{ioffe2015batch}.
As a comparison, Fig.~\ref{fig:fig8}b shows t-SNE plots of the same features with TTIN applied.
Figure~\ref{fig:fig8}c shows representative multicontrast slices from each test site, including full multicontrast images~(first row), as well as cases with missing FLAIR~(second row) and contrast variations~(third row).
Separated clusters from each test site were observed in both shallow and deep layers in Fig.~\ref{fig:fig8}a, indicating mismatched features across different test sites.
However, this mismatch was not observed after applying TTIN, as shown in Fig.~\ref{fig:fig8}b.

\subsection{Robustness of $\tau_1$ and $\tau_2$ Parameters in the ISBI Test}
\label{tau12}
\begin{figure*}[!t]

\begin{minipage}[b]{1.0\linewidth}
  \centering
  \centerline{\includegraphics[width=12cm]{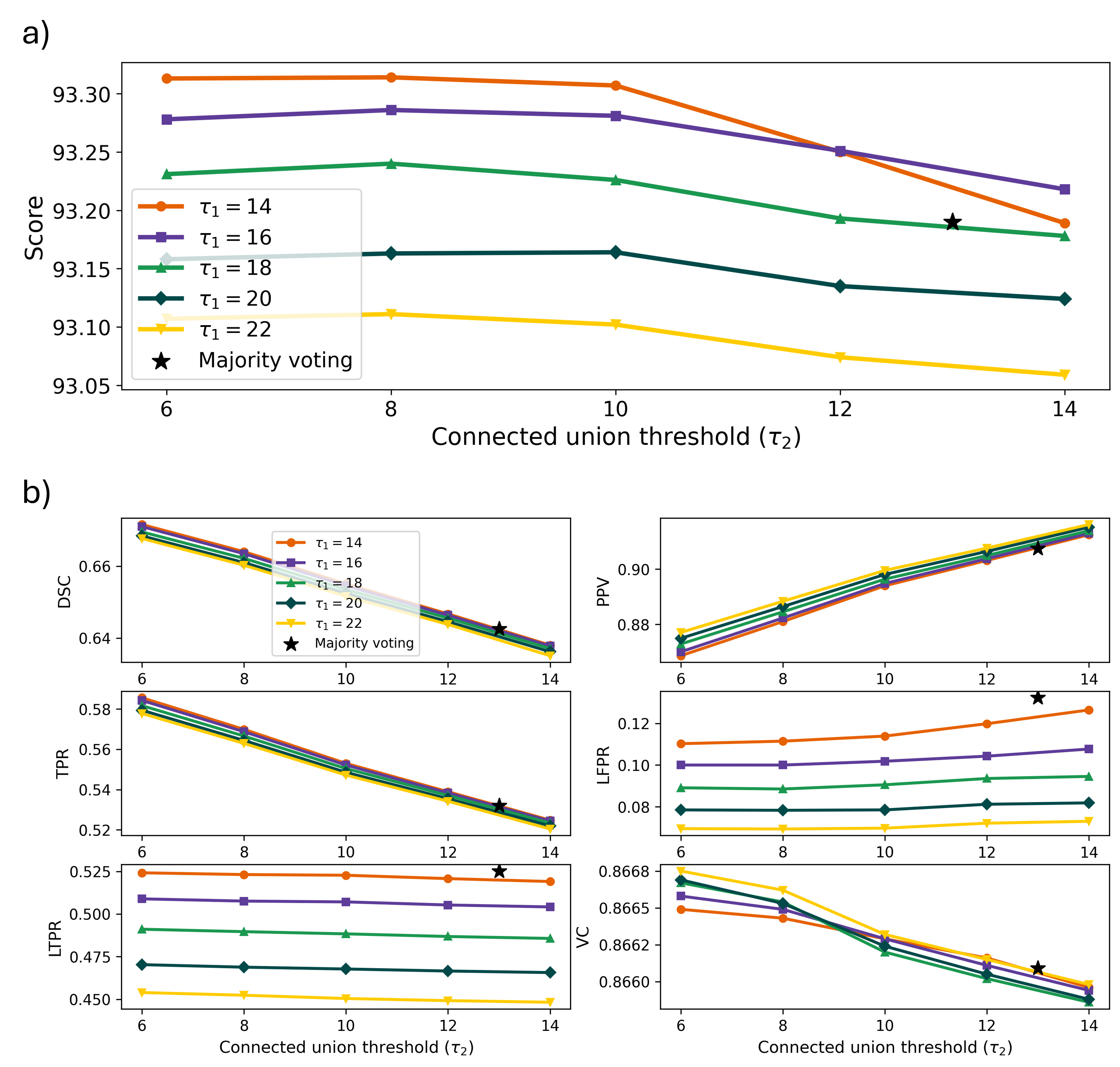}}
\end{minipage}

\caption{
(a): The robustness of the lesion detection $\tau_1$ (in Eq.~\eqref{voting1}) and connected lesion growth $\tau_2$ (in Eq.~\eqref{voting2}) parameters on the segmentation score~(Eq.~\eqref{score}) in the ISBI challenge test.
(b): The impact of $\tau_1$ and $\tau_2$ on individual metrics. 
For comparison, segmentation metrics from majority voting ($\tau_1 = \tau_2 = 13$, marked with stars) are also presented. 
}
\label{fig:fig9}
\end{figure*}

\textcolor{black}{In addition to the main test experiments in Sections~4.4 and 4.5 using the finalized UNISELF, we conducted separate comparative analyses in Sections~\ref{tau12} thru~\ref{ttin_outofdomin} on both in-domain and out-of-domain test datasets, evaluating different configurations of the self-ensembled lesion fusion (Section~~\ref{ssec:self}) and TTIN (Section~\ref{ssec:ttin}) components.
These analyses were motivated by the consistently strong cross-validation performance (Fig.~\hyperlink{figS1}{S1}) observed across varying \(\tau_1\) and \(\tau_2\) values (Eqs.~\ref{voting1} and~\ref{voting2}) and different normalization strategies for learning \(\gamma_j\) and \(\beta_j\) (Eq.~\ref{normalization_b}).}
A robustness analysis of the lesion detection $\tau_1$ (in Eq.~\eqref{voting1}) and connected lesion growth $\tau_2$ (in Eq.~\eqref{voting2}) parameters was conducted to evaluate their impact on segmentation performance using the ISBI challenge test dataset.
Figure~\ref{fig:fig9} illustrates the impact of $\tau_1$ and $\tau_2$ on the overall segmentation score (Eq.~\eqref{score}) and on individual metrics.
In Fig.~\ref{fig:fig9}a, we observe that an optimal score of $93.32$ was achieved with $\tau_1=14 \text{ and } \tau_2=8$.
Consistent scores above $93.20$ were achieved for $14 \leq \tau_1 \leq 18$ and $6 \leq \tau_2 \leq 12$, outperforming the majority voting~($\tau_1=\tau_2=13$) score of 93.19.
Figure~\ref{fig:fig9}b shows the effect of $\tau_1\text{ and } \tau_2$ on individual metrics.
In Fig.~\ref{fig:fig9}b, we observe that the change in $\tau_1$ at each fixed $\tau_2$ has relatively little impact on DSC, PPV and TPR, while larger values of $\tau_2$ contribute to improved PPV at the expense of decreased DSC and TPR.
However, LFPR and LTPR were primarily affected by $\tau_1$, with higher values of $\tau_1$ resulting in improved LFPR but worse LTPR.
Notably, only very small variations in VC are observed across different $\tau_1$ and $\tau_2$ values.

\subsection{Impact of TTIN on CD Performance in the ISBI Test}
\label{ttin_indomin}
For the ISBI test dataset, we conducted a comprehensive experiment using all input contrast combinations and investigated the impact of TTIN on models trained with and without CD. 
In this experiment, TTIN was compared with the use of BN training dataset statistics for feature normalization at test time, as originally proposed in BN~\citep{ioffe2015batch}.
TTIN with various feature normalization methods for training $\gamma_j$ and $\beta_j$ in Eq.~\eqref{normalization_b} was also employed and compared, including BN, IN, and CondIN as described in Section~\ref{ssec:ttin}.
The proposed self-ensemble in Section~\ref{ssec:self} was applied to all cases.
\textcolor{black}{Statistical significance~($p<0.05$) was assessed using the paired Wilcoxon signed-rank test over 122 subject-level average scores~(61 test subjects evaluated against two expert rater references), with each score computed by averaging performance across all 15 input contrast combinations.}

First, our results~(see Table~\hyperlink{tableS1}{S1}) show that the performance was consistently higher in independently trained models without CD for each input contrast combination compared to those with CD, especially when the FLAIR contrast was removed from the input.
Second, for models trained with BN, switching from BN training statistics to TTIN statistics at test time improved the performance of CD-trained models~\textcolor{black}{($p<0.05$ comparing the second and fourth columns in Table~\hyperlink{tableS1}{S1})}.
Specifically, this resulted in a noticeable improvement in test scores with TTIN statistics (fourth column in Table~\hyperlink{tableS1}{S1}) compared to BN training statistics (second column in Table~\hyperlink{tableS1}{S1}) \textcolor{black}{for input contrast combinations~(rows) highlighted by a teal-to-gold color-shaded value change.}
Furthermore, no significant differences were observed among CD-trained TTIN models using BN, CondIN, and IN~\textcolor{black}{($p>0.05$ for all paired comparisons among the three models)}.
For complete segmentation scores in this experiment, see Table~\hyperlink{tableS1}{S1}.

\subsection{Impact of TTIN on CD Performance in Out-of-Domain Tests}
\label{ttin_outofdomin}

\begin{figure*}[!t]

\begin{minipage}[b]{1.0\linewidth}
  \centering
  \centerline{\includegraphics[width=18cm]{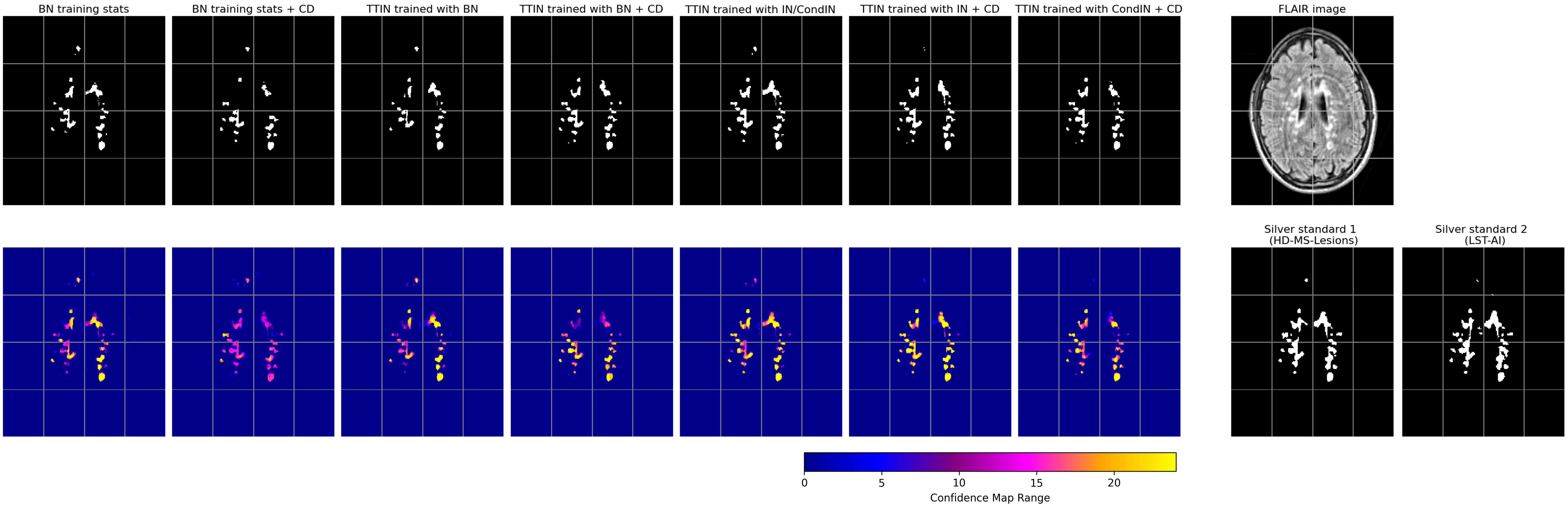}}
\end{minipage}
\caption{
Impact of TTIN (Section~\ref{ssec:ttin}) on segmentation masks comparing models trained with and without contrast dropout~(CD).
The same subject with FLAIR motion artifacts as depicted in Fig.~\ref{fig:figS2} is presented.
}
\label{fig:fig10}
\end{figure*}

We also conducted experiments to investigate the impact of TTIN on models trained with and without CD using Test Datasets~\RNum{2}, \RNum{3}, and~\RNum{4}.
The proposed self-ensemble strategy~(Section~\ref{ssec:self}) was applied to all models.
Results~(see Tables~\hyperlink{tableS2}{S2} and~\hyperlink{tableS3}{S3}) show that when TTIN was employed, CD-trained models consistently outperformed independently trained models without CD across all test datasets.
However, when BN training statistics were employed, CD-trained models showed degraded performance compared to independently trained models when the FLAIR contrast was missing.
Comparable and optimal performance was observed on UNISELF trained with BN, IN, and CondIN.
For the complete segmentation scores
with statistical significance tests in these experiments, please refer to Tables~\hyperlink{tableS2}{S2} and~\hyperlink{tableS3}{S3}.

Figure~\ref{fig:fig10} shows segmentation results and their corresponding confidence maps (Eq.~\eqref{confidence_map}) for the methods listed in Table~\hyperlink{tableS3}{S3} on a representative subject from the private test dataset with FLAIR motion artifacts.
Models using BN training statistics, both with and without CD, exhibited under-segmentation of lesions (columns 1-2).
Moreover, TTIN trained with BN but without CD also exhibited noticeable under-segmentation of lesions (column 3).
Applying CD resolved the under-segmentation issue (column 4).
Similarly, no noticeable missing lesions were observed in TTIN trained with IN and CondIN when CD was applied (columns 6-7).

\section{Discussion}
\label{sec:discussion}

\subsection{Self-Ensembled Lesion Fusion Improves Segmentation Accuracy}
UNISELF employs a novel self-ensembled lesion fusion strategy~(Section~\ref{ssec:self}) for TTA to improve segmentation accuracy.
\textcolor{black}{The proposed two-step lesion fusion strategy outperformed the conventional single-step voting approaches~(last two rows in Table~\ref{tab:table1}) commonly used in prior methods~\citep{isensee2021nnu, henschel2020fastsurfer}.}
In UNISELF, the self-ensemble parameters $\tau_1$ and $\tau_2$ primarily affect \textcolor{black}{lesion-wise and voxel-wise metrics}, respectively~(Fig.~\ref{fig:fig9}b).  
A possible explanation is that voxel-wise metrics such as DSC, PPV, and TPR are more susceptible to the total volume of segmented lesions, primarily influenced by $\tau_2$.
In contrast, lesion-wise metrics such as LFPR and LTPR are largely determined by lesion detection, which $\tau_1$ mainly affects.
This concept of ``detection and growth'' using high and low threshold values was also proposed and proven effective in Canny edge detection~\citep{canny1986computational} with Gaussian-filtered edge detection to generate candidate edges before thresholding.
Compared to Canny edge detection, UNISELF generates candidate lesions ($C(\r)$ in Fig.~\ref{fig:fig2}a) through its comprehensive multi-orientation processing of 3D multicontrast MR volumes~(Fig.~\ref{fig:fig1}) using a unified segmentation model.

\subsection{TTIN Enhances Generalization to Varying Contrasts}
UNISELF leverages TTIN~(Section~\ref{ssec:ttin}) to improve \textcolor{black}{model generalization handling missing contrasts}.
Its effectiveness \textcolor{black}{in addressing varying contrast combinations in CD-trained models} was validated using the ISBI challenge test data~(Table~\hyperlink{tableS1}{S1}).
Simply switching to TTIN stats after BN training, as indicated by `TTIN trained with BN' in Table~\hyperlink{tableS1}{S1}, significantly improves the performance of CD-trained models.
This indicates that when training a lesion segmentation model with CD, sharing BN feature statistics among contrast combinations is not sufficient for optimal performance across all combinations. 
There is a tendency for these statistics to favor combinations that include FLAIR, thus sacrificing the performance of combinations without FLAIR.
In contrast, TTIN trained with BN, IN, and CondIN avoid this issue by applying instance-specific feature normalization to each input combination, thereby achieving performance comparable to their independently trained counterparts.
\textcolor{black}{Unlike imputation-based methods, which attempt to synthesize missing contrasts through image translation models that often require complex training pipelines and risk hallucinated or low-fidelity content, UNISELF addresses missing contrast without explicit synthesis. 
Specifically, CD during training exposes the model to random contrast combinations, while TTIN recalibrates feature distributions at inference by updating normalization statistics for each contrast combination.
This design enables effective generalization to real-world scenarios where ideal contrast combinations may not be available.}

\subsection{Validation on Out-of-Domain and Missing FLAIR Scenarios}
The impact of TTIN on model generalization was further validated using various out-of-domain test datasets~(Tables~\hyperlink{tableS2}{S2} and~\hyperlink{tableS3}{S3}).
For both CD-trained and independently trained models, inferior performance was observed in models using BN training stats compared to those using TTIN stats.
This inferiority was more pronounced for CD-trained models when the FLAIR contrast was missing (`No FLAIR' in Table~\hyperlink{tableS2}{S2}).
In fact, CD training has been reported to improve domain generalization~\citep{zhang2023domain}. 
Our results support this finding, with an important contribution that demonstrates the need to employ TTIN for the effective implementation of CD training.
\textcolor{black}{FLAIR is well known to be an ideal contrast for MS diagnosis and monitoring~\citep{wattjes2015evidence} and our missing FLAIR experiments also support that position.
Unfortunately the missing FLAIR scenario commonly arises in multi-center or longitudinal studies involving legacy data, where FLAIR may be missing due to historical protocol choices or inconsistent acquisition.
Additionally, in non-MS neurological disorder cohorts (e.g., ADNI 1 and 2~\citep{jack2008alzheimer}), where MRI is primarily used for general clinical assessment, FLAIR is often not prioritized and may be entirely absent. 
However, white matter lesion segmentation may still be required, making model reliability under missing FLAIR conditions highly desirable.}

\subsection{Comparison with Other Test-Time Adaptation Strategies}
\textcolor{black}{Prior work has explored test-time strategies to address distribution shifts.
\citet{nado2020evaluating} demonstrated that recomputing BN statistics at prediction time for each test batch improves model robustness under domain shift.
\citet{gerin2024exploring} applied test-time training (TTT) with auxiliary tasks on a per-patient basis to mitigate domain shifts without requiring data from other patients.
In comparison, we adopt TTIN in UNISELF, which offers several advantages for MS lesion segmentation.
First, unlike prediction-time BN, TTIN operates per test input without batch size dependency, making it directly applicable on-the-fly in clinical workflows.
Second, unlike TTT, TTIN is lightweight and training-free, requiring no additional model updates.
Third and most importantly, we show that TTIN is especially beneficial for CD-trained models, as it mitigates intrinsic distribution shifts from varying contrast combinations even when the model is exposed to all combinations during training.
Our work complements recent benchmarking efforts on domain shift~\citep{malinin2022shifts}, demonstrating that TTIN offers a practical and robust alternative to TTT and prediction-time BN in MS lesion segmentation.}

\subsection{Generalization without Large-Scale Annotated Data}
\textcolor{black}{State-of-the-art MS lesion segmentation methods trained on large and heterogeneous datasets, such as HD-MS-Lesions~\citep{brugnara2020automated} and LST-AI~\citep{wiltgen2024lst}, represent another strategy for improving model generalization. 
These methods benefit from broad data diversity and extensive expert annotations, which help capture inter-site and inter-scanner variations.
However, such large-scale annotated datasets are often not publicly available.
In contrast, our method is designed for scenarios where access to large heterogeneous datasets is not feasible and where protocol variability and missing contrasts are major concerns. 
By leveraging self-ensembled lesion fusion and TTIN, UNISELF achieves strong generalization performance under limited training data and in the presence of domain shifts or missing MRI contrasts.}

\subsection{Broader Applications}
In addition to MS lesion segmentation, there is growing interest in developing automated detection methods for paramagnetic rim lesions in MS~\citep{zhang2022qsmrim, barquero2020rimnet, lou2021fully}.
However, the generalization ability of these methods for multi-site deployment has not been fully validated.
As future work, the proposed self-ensemble and test-time instance normalization in UNISELF may also be applied to rim lesion detection, \textcolor{black}{for example, by aggregating predictions across multi-orientation inputs using a tunable threshold similar to $\tau_1$ in UNISELF}, with rigorous multi-site validation.
Beyond downstream MS segmentation and rim lesion detection, there has been active research in developing cross-domain generalizable deep learning models for low-level image formation tasks with limited training data.
For instance, there has been increased interest in enhancing the generalization capacity of MRI reconstruction models trained on single-vendor data, particularly against deviations in acquisition parameters and vendor changes~\citep{muckley2021results, zhang2023laro}. 
This improvement is facilitated by incorporating a self-adaptive k-space data consistency module within the networks.
Other examples include test-time domain adaptation to unseen domains, which is achieved using data consistency loss~\citep{zhang2020fidelity, zhang2020probabilistic, yoo2021time, gungor2023adaptive}.
UNISELF for MS lesion segmentation resembles self-adaptive data consistency for MRI reconstruction in terms of its full exploitation of the input data through self-ensemble and self-adaptive latent features through TTIN.
\textcolor{black}{Notably, self-ensemble can also be applied to MRI reconstruction by averaging reconstructions from rotated and flipped k-space data and leveraging their variance to assess reconstruction uncertainty.}

\subsection{Limitations and Improvement}
There are some limitations of this work.
First, the two thresholds $\tau_1$ and $\tau_2$ are fixed across all locations in the brain, which may not be optimal for detecting and segmenting certain types of lesions in MS, such as small cortical lesions~\citep{la2020multiple}.
Implementing spatially adaptive thresholds with lower $\tau_1$ and $\tau_2$ values near the cortical regions may improve the sensitivity of cortical lesion detection and segmentation.
Second, although processing multi-orientation inputs using a single network~(Fig.~\ref{fig:fig1}) is advantageous for implementation simplicity in UNISELF, exploring advanced architectures for more effective multi-orientation processing is still needed.
The Mixture-of-Experts (MoE) architecture~\citep{jacobs1991adaptive} is an option that treats multi-orientation inputs as subproblems to form an expert-guided pathway for a given input data.
Third, although the silver standard delineations generated for the private multi-site dataset were carefully crafted, they were still not as convincing as the expert delineations available in other test datasets.
Finally, while UNISELF demonstrated good generalization when trained with limited single-site labeled data, its generalization may be further improved by leveraging large-scale unlabeled clinical multicontrast MRI datasets through self-supervised pretraining~\citep{tang2022self} or by training with synthetic MRI containing MS lesions~\citep{zhang2024bi}. 

\section{Conclusion}
\label{sec:conclusion}
We propose UNISELF, a state-of-the-art MS lesion segmentation method that integrates a novel self-ensembled lesion fusion strategy and leverages test-time instance normalization for improved accuracy and generalization.
Trained on the 2015 ISBI challenge training dataset, UNISELF ranks among the best performing methods on the challenge's test dataset.
Additionally, UNISLEF has demonstrated better generalization than the benchmark methods to out-of-domain datasets with contrast variation, imaging artifacts, and missing contrast.

\section*{Acknowledgments}
This work was partially supported by the National Science Foundation Graduate Research Fellowship under Grant Nos. DGE-1746891~(S.~W.~Remedios) and DGE-2139757~(S.~P.~Hays).
This work was also supported in part by the Department of Defense in the Center for Neuroscience and Regenerative Medicine and by the NIH through the Intramural Research Program of NINDS and NINDS grants R01-NS082347 (PI: P.~A. Calabresi) and U01-NS111678 (PI: P.~A. Calabresi).
It was also supported by the Congressionally Directed Medical Research Programs grants W81XWH2010912 (PI: J.~L.~Prince), HT94252410785 (PI: B.~E.~Dewey), and HT94252510716 (PI: J.~Zhang); the National MS Society grants RG-1507-05243 (PI: D.~L.~Pham) and FG-2008-36966 (PI: B.~E.~Dewey); and the PCORI grant MS-1610-37115 (PIs: S.~D.~Newsome and E.~M.~Mowry), which also provided support for J.~Zhang, L.~Zuo, B.~E.~Dewey, S.~W.~Remedios, S.~P.~Hays, and A.~Carass.
L.~Zuo was supported by the NCI grants R01-CA253923 and R01-CA275015.
Y.~Liu was supported by the NIH grants R01HL169944, U24AG074855, and R01MH121620.
The statements in this publication are solely the responsibility of the authors and do not necessarily represent the views of PCORI, its Board of Governors or Methodology Committee.

\bibliographystyle{model2-names.bst}\biboptions{authoryear}
\bibliography{refs}

\clearpage
\renewcommand\thefigure{S\arabic{figure}}
\setcounter{figure}{0}
\renewcommand\thetable{S\arabic{table}}
\setcounter{table}{0}
\section*{Supplementary Material}\nopagebreak[4]

\nopagebreak[4]
%
\begin{table*}[!h]
\centering
\hypertarget{tableS1}{}
\caption{Impact of TTIN (Section~\ref{ssec:ttin}) on segmentation scores~(Eq.~\eqref{score}) comparing models trained with and without contrast dropout (CD).
When CD was not applied, independent models were trained separately for each input contrast combination, where CondIN and IN are identical for independently trained models.
All input multicontrast image combinations from the 2015 ISBI challenge test dataset were tested and compared.
Self-ensembled lesion fusion (Section~\ref{ssec:self}) was applied to all models.
\textcolor{black}{
Rows highlighted in teal and gold indicate input contrast combinations where the same CD-trained model shows a noticeable improvement when switching from BN training statistics to TTIN statistics at test time.}
}
\label{tab:tableS1}

\rowcolors{3}{white}{gray!35}

\scalebox{0.65}{
    \begin{tabular}{
            cccc l cc l | cc l cc l cc
    }
    \toprule
    \multicolumn{4}{c}{\textbf{Input contrasts}} && \multicolumn{2}{c}{\textbf{BN training stats}} && \multicolumn{2}{c}{\textbf{TTIN trained with BN}} && \multicolumn{2}{c}{\textbf{TTIN trained with IN}} &&
    \multicolumn{2}{c}{\textbf{TTIN trained with CondIN}} \\
    \cmidrule(lr){1-4}
    \cmidrule(lr){6-16}
        \rowcolor{white}\textbf{T1w} & \textbf{T2w} & \textbf{PDw} & \textbf{FLAIR}  && \textbf{without CD} & \textbf{with CD} && \textbf{without CD} & \textbf{with CD} && \textbf{without CD} & \textbf{with CD} && \textbf{without CD} & \textbf{with CD (Proposed)} \\
    \cmidrule(lr){1-4}
    \cmidrule(lr){6-16}
    
$\blacktriangle$ & $\blacktriangle$ & $\blacktriangle$ & $\blacktriangle$ & & 93.204 ± 7.351 & 93.134 ± 7.070 & & 93.174 ± 8.383 & 93.078 ± 7.613 & & 93.350 ± 6.583 & 93.271 ± 7.234 & & 93.350 ± 6.583 & 93.286 ± 7.400 \\ 
$\blacktriangle$ & $\blacktriangle$ & $\blacktriangle$ &  & & 92.532 ± 6.886 & \cellcolor{lowteal}92.093 ± 7.142 & & 92.483 ± 7.762 & \cellcolor{highgold}92.319 ± 7.286 & & 92.620 ± 7.419 & 92.353 ± 7.435 & & 92.620 ± 7.419 & 92.386 ± 7.238 \\ 
$\blacktriangle$ & $\blacktriangle$ &  & $\blacktriangle$ & & 93.255 ± 6.764 & 93.039 ± 7.110 & & 93.171 ± 7.774 & 93.112 ± 7.721 & & 93.319 ± 6.958 & 93.129 ± 7.645 & & 93.319 ± 6.958 & 93.172 ± 7.933 \\ 
$\blacktriangle$ &  & $\blacktriangle$ & $\blacktriangle$ & & 93.121 ± 7.662 & 93.028 ± 7.541 & & 93.086 ± 7.763 & 93.074 ± 7.801 & & 93.184 ± 7.227 & 93.036 ± 7.743 & & 93.184 ± 7.227 & 93.145 ± 7.686 \\ 
 & $\blacktriangle$ & $\blacktriangle$ & $\blacktriangle$ & & 93.253 ± 7.084 & \cellcolor{lowteal}92.494 ± 8.448 & & 93.017 ± 7.736 & \cellcolor{highgold}92.986 ± 7.782 & & 93.179 ± 7.508 & 93.117 ± 7.300 & & 93.179 ± 7.508 & 93.131 ± 7.554 \\ 
$\blacktriangle$ & $\blacktriangle$ &  &  & & 91.608 ± 6.830 & \cellcolor{lowteal}88.714 ± 5.230 & & 91.694 ± 6.652 & \cellcolor{highgold}91.777 ± 7.056 & & 91.800 ± 7.688 & 91.981 ± 6.924 & & 91.800 ± 7.688 & 91.891 ± 7.326 \\ 
$\blacktriangle$ &  & $\blacktriangle$ &  & & 92.035 ± 6.032 & \cellcolor{lowteal}91.221 ± 7.098 & & 92.209 ± 6.918 & \cellcolor{highgold}91.839 ± 7.862 & & 92.248 ± 6.768 & 92.028 ± 7.169 & & 92.248 ± 6.768 & 91.987 ± 7.001 \\ 
$\blacktriangle$ &  &  & $\blacktriangle$ & & 93.114 ± 7.856 & \cellcolor{lowteal}92.473 ± 7.642 & & 93.099 ± 7.635 & \cellcolor{highgold}92.768 ± 8.439 & & 93.089 ± 7.272 & 92.905 ± 7.980 & & 93.089 ± 7.272 & 92.987 ± 7.993 \\ 
 & $\blacktriangle$ & $\blacktriangle$ &  & & 92.293 ± 7.539 & 91.820 ± 6.387 & & 92.374 ± 8.312 & 91.914 ± 7.342 & & 92.483 ± 7.919 & 92.307 ± 7.834 & & 92.483 ± 7.919 & 92.414 ± 7.132 \\ 
 & $\blacktriangle$ &  & $\blacktriangle$ & & 93.249 ± 7.417 & \cellcolor{lowteal}91.714 ± 7.675 & & 93.048 ± 7.866 & \cellcolor{highgold}92.701 ± 8.387 & & 93.137 ± 7.781 & 92.905 ± 7.871 & & 93.137 ± 7.781 & 92.870 ± 8.187 \\ 
 &  & $\blacktriangle$ & $\blacktriangle$ & & 92.985 ± 7.978 & \cellcolor{lowteal}90.739 ± 12.58 & & 92.794 ± 8.260 & \cellcolor{highgold}92.609 ± 8.653 & & 93.002 ± 8.124 & 92.866 ± 7.904 & & 93.002 ± 8.124 & 92.867 ± 7.989 \\ 
$\blacktriangle$ &  &  &  & & 91.117 ± 6.226 & \cellcolor{lowteal}nan ± nan & & 91.078 ± 6.520 & \cellcolor{highgold}90.786 ± 6.645 & & 91.264 ± 6.577 & 91.070 ± 6.674 & & 91.264 ± 6.577 & 91.053 ± 6.614 \\ 
 & $\blacktriangle$ &  &  & & 90.420 ± 11.23 & \cellcolor{lowteal}85.240 ± 39.69 & & 91.408 ± 7.565 & \cellcolor{highgold}90.972 ± 5.896 & & 91.470 ± 6.849 & 91.541 ± 6.958 & & 91.470 ± 6.849 & 91.471 ± 7.307 \\ 
 &  & $\blacktriangle$ &  & & 90.494 ± 9.037 & \cellcolor{lowteal}87.766 ± 16.06 & & 91.494 ± 6.865 & \cellcolor{highgold}90.818 ± 7.920 & & 91.628 ± 7.350 & 91.187 ± 7.402 & & 91.628 ± 7.350 & 91.129 ± 7.551 \\ 
 &  &  & $\blacktriangle$ & & 92.256 ± 9.322 & \cellcolor{lowteal}85.727 ± 38.16 & & 92.089 ± 9.331 & \cellcolor{highgold}91.684 ± 10.81 & & 92.230 ± 9.106 & 92.184 ± 9.340 & & 92.230 ± 9.106 & 92.070 ± 9.465 \\
    \bottomrule
    \end{tabular}
}
\end{table*}

\begin{table*}[!h]
\centering
\hypertarget{tableS2}{}
\caption{
Impact of TTIN (Section~\ref{ssec:ttin}) on segmentation scores (Eq.~\eqref{score}) comparing models trained with and without contrast dropout (CD).
When CD was not applied, independent models were trained separately for each input contrast combination.
Public and private test datasets considering both original and missing FLAIR multicontrast inputs were tested and compared.
Self-ensemble (Section~\ref{ssec:self}) were applied to all models.
}
\label{tab:tableS2}

\rowcolors{3}{white}{gray!35}

\scalebox{0.7}{
\begin{tabular}{
        cc l cc l cc l cc
}
\toprule
\rowcolor{white} 
\multicolumn{2}{c}{\textbf{Method}} && 
\multicolumn{2}{c}{\textbf{2016 MICCAI Dataset}} && 
\multicolumn{2}{c}{\textbf{UMCL Dataset}} &&
\multicolumn{2}{c}{\textbf{Private Dataset}} \\
\cmidrule(lr){1-2}
\cmidrule(lr){3-11}
\rowcolor{white} 
\textbf{Norm Stats} & \textbf{CD}  && 
\textbf{Original} & \textbf{No FLAIR} && 
\textbf{Original} & \textbf{No FLAIR} && 
\textbf{Original} & \textbf{No FLAIR} \\
\rowcolor{white} 
& && 
\textbf{(T1+T2+PD+FLAIR)} & \textbf{(T1+T2+PD)} && 
\textbf{(T1+T2+FLAIR)} & \textbf{(T1+T2)} && 
\textbf{(mixed inputs)} & \\
\cmidrule(lr){1-2}
\cmidrule(lr){3-11}

BN training stats & \xmark & & \pStar{}0.746 ± 0.062\bStar{} & 0.711 ± 0.039 && \pStar{}0.720 ± 0.098\bStar{} & \pStar{}0.651 ± 0.088\bStar{} && \pStar{}0.744 ± 0.099\bStar{} & \pStar{}0.679 ± 0.092\bStar{} \\
BN training stats & \cmark & & \bf{0.773 ± 0.067} & \bf{0.723 ± 0.064} && \pStar{}0.731 ± 0.078\bStar{} & \pStar{}0.548 ± 0.065\bStar{} && \pStar{}0.732 ± 0.106\bStar{} & \pStar{}0.644 ± 0.120\bStar{} \\
\hline
TTIN trained with BN & \xmark & & \pStar{}0.734 ± 0.063\bStar{} & 0.709 ± 0.067 && \pStar{}0.724 ± 0.091\bStar{} & \pStar{}0.649 ± 0.113\bStar{} && \pStar{}0.758 ± 0.090\bStar{} & \pStar{}0.692 ± 0.088\bStar{} \\
TTIN trained with BN & \cmark & & 0.767 ± 0.062 & 0.711 ± 0.057 && \pStar{}0.730 ± 0.081\bStar{} & \underline{0.674 ± 0.065} && \bf{0.785 ± 0.078} & 0.720 ± 0.094 \\
TTIN trained with IN/CondIN & \xmark & & 0.753 ± 0.056 & 0.717 ± 0.065 && \underline{0.737 ± 0.103} & \pStar{}0.657 ± 0.097\bStar{} && \pStar{}0.759 ± 0.086\bStar{} & \pStar{}0.693 ± 0.097\bStar{} \\
TTIN trained with IN & \cmark & & 0.769 ± 0.060 & \underline{0.722 ± 0.058} && \bf{0.745 ± 0.090} & \bf{0.676 ± 0.077} && 0.781 ± 0.080 & \bf{0.726 ± 0.085} \\
TTIN trained with CondIN (Proposed) & \cmark & & \underline{0.770 ± 0.057} & 0.715 ± 0.066 && \bf{0.745 ± 0.085} & 0.668 ± 0.100 && \underline{0.783 ± 0.074} & \underline{0.722 ± 0.086} \\
\bottomrule

\end{tabular}
}
{\\ \footnotesize (\bStar{}: statistically significant in the Wilcoxon signed-rank test compared to `TTIN trained with CondIN + CD (Proposed)' (last row in the table) in each column, p-value $<$ 0.05.
The best and second-best performances in each column are denoted in \textbf{bold} and \underline{underline}, respectively.)}
\end{table*}
\begin{table*}[!h]
\centering
\hypertarget{tableS3}{}
\caption{
Same segmentation score comparison as Table~\ref{tab:tableS2} on the private test dataset with various FLAIR artifacts.
}
\label{tab:tableS3}

\rowcolors{3}{white}{gray!35}

\scalebox{0.7}{
\begin{tabular}{
        cc l cccccc
}
\toprule
\rowcolor{white} 
\multicolumn{2}{c}{Method} && 
\multicolumn{6}{c}{\textbf{FLAIR Artifacts in Private Dataset}} \\
\cmidrule(lr){1-2}
\cmidrule(lr){3-9}
\rowcolor{white} 
\textbf{Norm Stats} & \textbf{CD} && 
\textbf{Motion} & \textbf{Noise} & \textbf{Ghosting} & \textbf{Bias Field} & 
\textbf{Spatial Blurriness} & \textbf{Anisotropy} \\
\cmidrule(lr){1-2}
\cmidrule(lr){3-9}

BN training stats & \xmark & & \pStar{}0.720 ± 0.110\bStar{} & \pStar{}0.740 ± 0.107\bStar{} & \pStar{}0.726 ± 0.102\bStar{} & \pStar{}0.729 ± 0.096\bStar{} & \pStar{}0.733 ± 0.094\bStar{} & \pStar{}0.701 ± 0.109\bStar{} \\
BN training stats & \cmark & & \pStar{}0.699 ± 0.112\bStar{} & \pStar{}0.712 ± 0.113\bStar{} & \pStar{}0.704 ± 0.111\bStar{} & \pStar{}0.741 ± 0.090\bStar{} & \pStar{}0.746 ± 0.090\bStar{} & \pStar{}0.719 ± 0.101\bStar{} \\
\hline
TTIN trained with BN & \xmark & & \pStar{}0.735 ± 0.101\bStar{} & \pStar{}0.756 ± 0.093\bStar{} & \pStar{}0.734 ± 0.100\bStar{} & \pStar{}0.727 ± 0.109\bStar{} & \pStar{}0.741 ± 0.096\bStar{} & \pStar{}0.721 ± 0.093\bStar{} \\
TTIN trained with BN & \cmark & & \bf{0.765 ± 0.089} & \bf{0.782 ± 0.080} & \bf{0.767 ± 0.081} & \bf{0.778 ± 0.083} & \underline{0.772 ± 0.086} & \bf{0.761 ± 0.083} \\
TTIN trained with IN/CondIN & \xmark & & \pStar{}0.738 ± 0.092\bStar{} & \pStar{}0.759 ± 0.087\bStar{} & \pStar{}0.739 ± 0.087\bStar{} & \pStar{}0.744 ± 0.094\bStar{} & \pStar{}0.727 ± 0.103\bStar{} & \pStar{}0.716 ± 0.093\bStar{} \\
TTIN trained with IN & \cmark & & \bf{0.765 ± 0.094} & 0.778 ± 0.083 & \bf{0.767 ± 0.085} & 0.767 ± 0.087 & 0.770 ± 0.089 & 0.757 ± 0.085 \\
TTIN trained with CondIN (Proposed) & \cmark & & \underline{0.763 ± 0.096} & \underline{0.779 ± 0.079} & \underline{0.765 ± 0.083} & \underline{0.773 ± 0.083} & \bf{0.775 ± 0.085} & \underline{0.758 ± 0.080} \\
\bottomrule

\end{tabular}
}
{\\ \footnotesize (\bStar{}: statistically significant in the Wilcoxon signed-rank test compared to ` TTIN trained with CondIN + CD (Proposed)' (last row in the table) in each column, p-value $<$ 0.05.
The best and second-best performances in each column are denoted in \textbf{bold} and \underline{underline}, respectively.)}
\end{table*}

\begin{figure*}[!tb]
\begin{minipage}[b]{1.0\linewidth}
  \centering
  \centerline{\includegraphics[width=18cm]{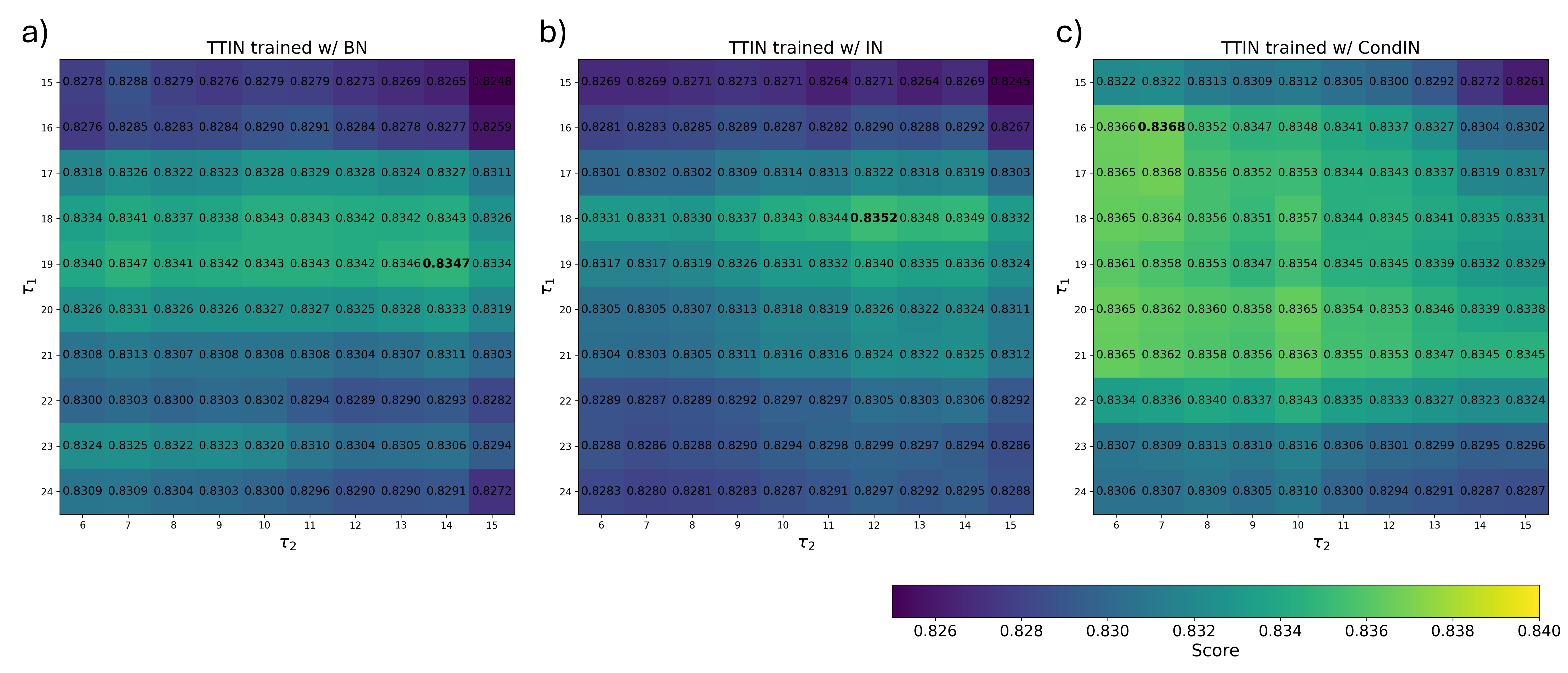}}
\end{minipage}
\hypertarget{figS1}{}
\caption{
    \textcolor{black}{Cross-validation grid search of self-ensembled lesion fusion parameters $\tau_1$ and $\tau_2$ for models employing TTIN trained with BN, IN, and CondIN. 
    Each heatmap shows the average validation score on the ISBI training set using five-fold cross-validation, with one subject held out for validation in each fold. 
    Scores are computed for each $(\tau_1, \tau_2)$ pair, and bolded values indicate the highest performance within each $(\tau_1, \tau_2)$ grid.}
    }
\label{fig:figS1}
\end{figure*}

\begin{figure*}[!tb]

\begin{minipage}[b]{1.0\linewidth}
  \centering
  \centerline{\includegraphics[width=18cm]{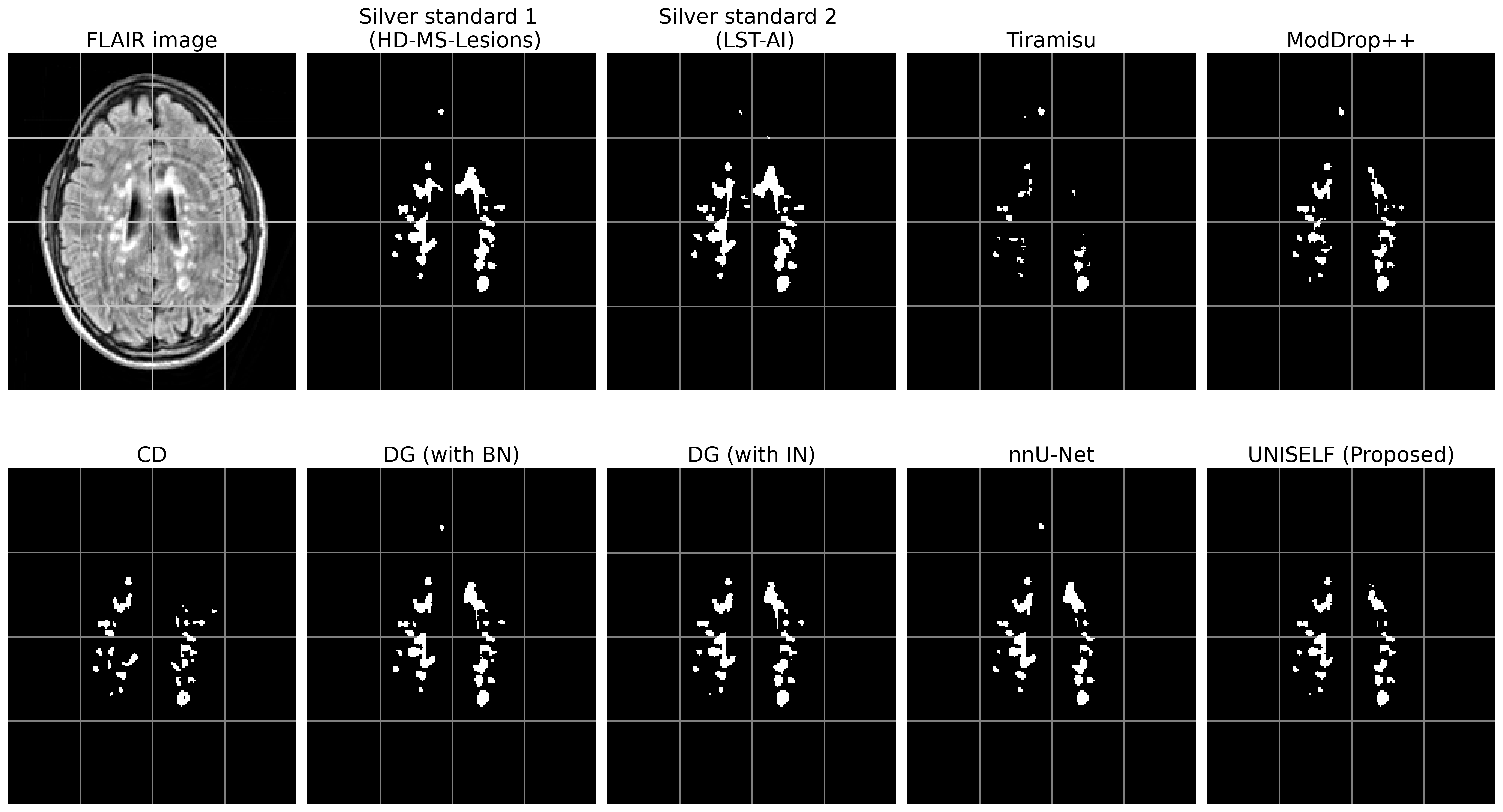}}
\end{minipage}
\hypertarget{figS2}{}
\caption{
MS segmentation masks from different methods on the same subject as in Fig.~\ref{fig:fig5}, with motion artifacts present on the FLAIR image. 
}
\label{fig:figS2}
\end{figure*}

\begin{figure*}[!tb]

\begin{minipage}[b]{1.0\linewidth}
  \centering
  \centerline{\includegraphics[width=18cm]{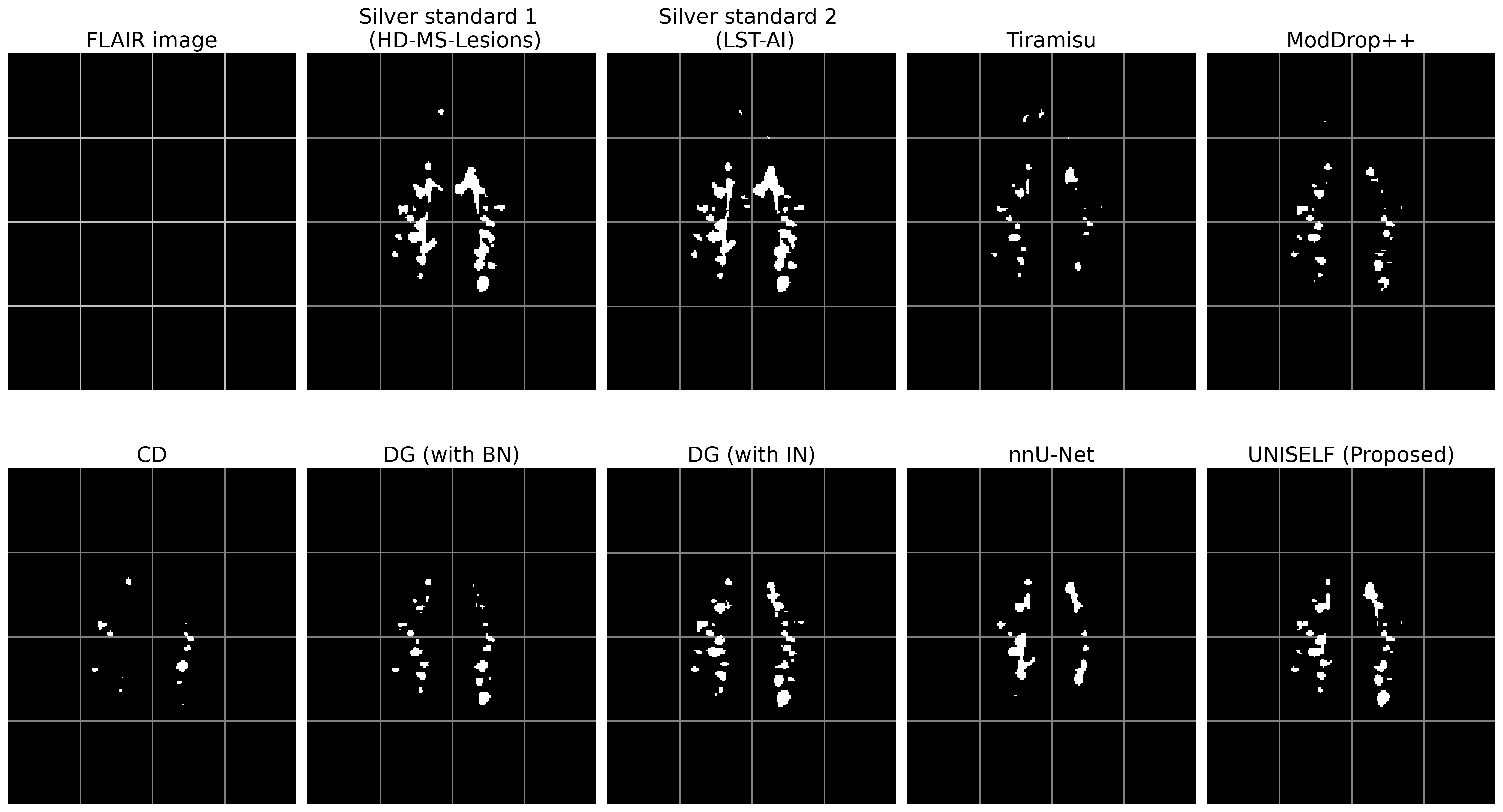}}
\end{minipage}
\hypertarget{figS3}{}
\caption{
MS segmentation masks from different methods on the same subject as in Figs.~\ref{fig:fig5} and~\ref{fig:figS2}, with missing FLAIR contrast.
}
\label{fig:figS3}
\end{figure*}

\begin{figure*}[!h]
\begin{minipage}[b]{1.0\linewidth}
  \centering
  \centerline{\includegraphics[width = \linewidth]{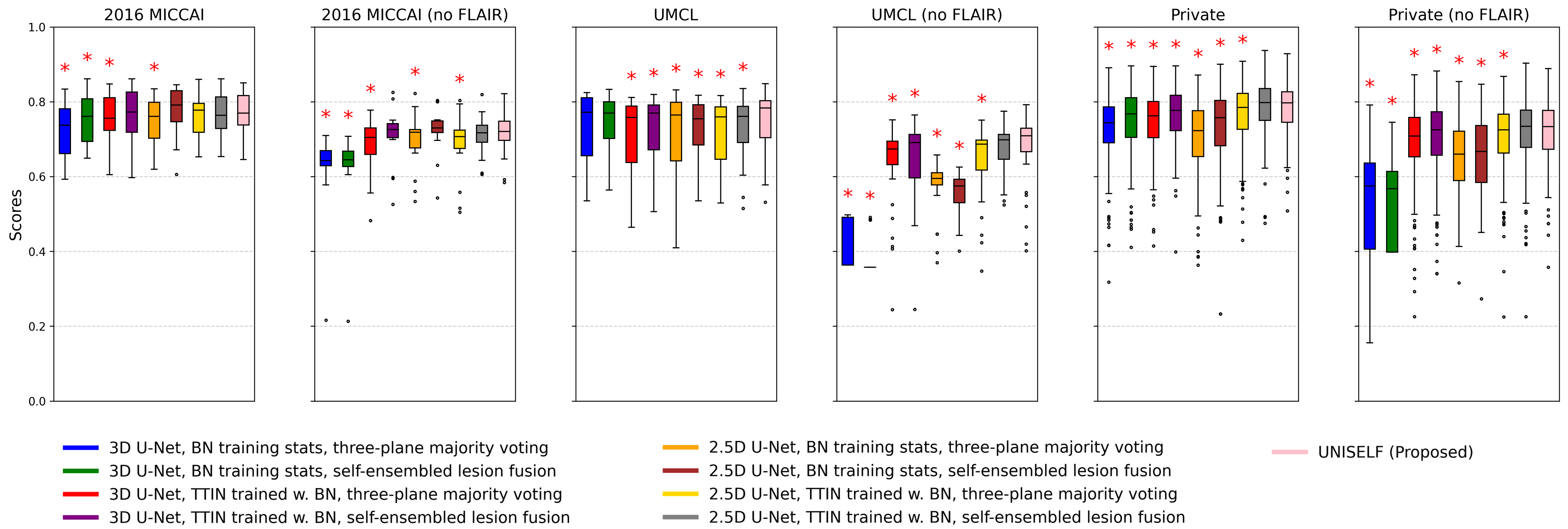}}
\end{minipage}
\hypertarget{figS4}{}
\caption{
    \textcolor{black}{
    Segmentation scores~(Eq.~\ref{score}) from the ablation study on public and private test datasets, considering both original and missing FLAIR multicontrast inputs.
    Both 2.5D~(as proposed) and 3D U-Net architectures were evaluated by progressively 1)~replacing ``self-ensembled lesion fusion'' with ``three-plane majority voting'' and 2)~replacing ``TTIN trained with BN'' with ``BN training stats''.
    ``UNISELF~(Proposed)'' refers to the UNISELF configuration finalized in cross-validation (Section~\ref{cross_val}), which uses TTIN trained with CondIN.
    (Red star: statistically significant difference compared to UNISELF (Proposed) in each boxplot, based on the paired Wilcoxon signed-rank test, $p < 0.05$.)
    }
}
\label{fig:figS4}
\end{figure*}

\begin{figure*}[!h]
\begin{minipage}[b]{1.0\linewidth}
  \centering
  \centerline{\includegraphics[width = \linewidth]{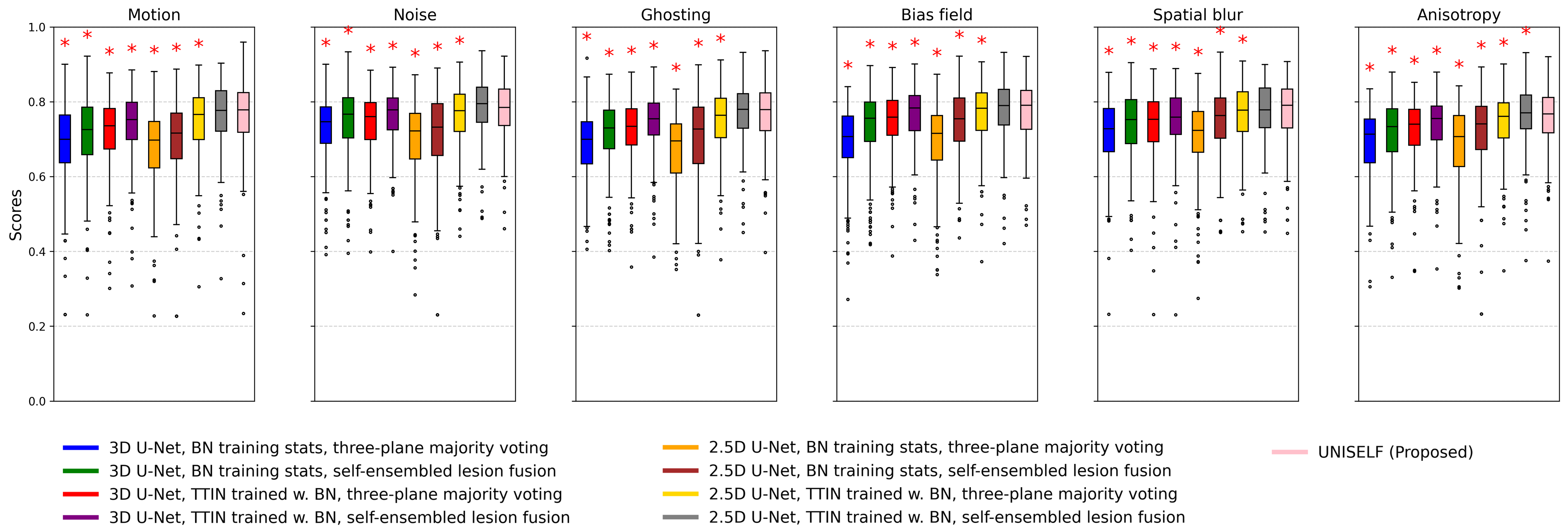}}
\end{minipage}
\hypertarget{figS5}{}
\caption{
    \textcolor{black}{
    Segmentation scores (Eq.~\ref{score}) from the ablation study on the private multisite test dataset containing various FLAIR artifacts.
    The same ablation procedure as in Fig.~\ref{fig:figS4} was applied.
    (Red star: statistically significant difference compared to UNISELF (Proposed) in each boxplot, based on the paired Wilcoxon signed-rank test, $p < 0.05$.)
    }
}
\label{fig:figS5}
\end{figure*}

\end{document}